\documentclass[hidelinks]{IEEEoj}
\usepackage[dvipsnames]{xcolor}

\usepackage[english]{babel}
\usepackage{blindtext}
\usepackage[utf8]{inputenc}
\usepackage{graphicx}
\usepackage{amsmath,amssymb,amsthm,mathtools}
\usepackage{upgreek}
\usepackage{ragged2e}
\usepackage{float}
\usepackage{bm}
\usepackage{cite}
\usepackage{cases}
\usepackage{subfigure}
\usepackage{hyperref}
\usepackage{mathrsfs}
\usepackage{multicol}
\usepackage{algorithm,algpseudocode}
\usepackage[acronym,shortcuts]{glossaries}

\usepackage{bm}
\usepackage{lipsum}
\usepackage{mleftright}
\usepackage{orcidlink}
\usepackage{booktabs}
\usepackage{fix-cm}
\usepackage{setspace}
\usepackage{xparse}
\usepackage[bottom]{footmisc}

\def\BibTeX{{\rm B\kern-.05em{\sc i\kern-.025em b}\kern-.08em
    T\kern-.1667em\lower.7ex\hbox{E}\kern-.125emX}}
\AtBeginDocument{\definecolor{tmlcncolor}{cmyk}{0.93,0.59,0.15,0.02}\definecolor{NavyBlue}{RGB}{0,86,125}}

\def\OJlogo{\vspace{-4pt}$<$Society logo(s) and publication title will appear here.$>$}

\def\authorrefmark#1{\ensuremath{^{\textbf{#1}}}}

\newacronym{ISAC}{ISAC}{integrated sensing and communications}
\newacronym{SVD}{SVD}{singular value decomposition}
\newacronym{DCM}{DCM}{double-centering matrix}
\newacronym{3D}{3D}{three-dimensional}
\newacronym{GA}{GA}{genie-aided}
\newacronym{EA}{EA}{``\emph{estimate-then-average}''}
\newacronym{AE}{AE}{``\emph{average-then-estimate}''}
\newacronym{IRS}{IRS}{intelligent reflecting surface}
\newacronym{RSSI}{RSSI}{received signal strength indicator}
\newacronym{SotA}{SotA}{state-of-the-art}
\newacronym{CSI}{CSI}{channel state information}
\newacronym{D2D}{D2D}{device-to-device}
\newacronym{RR}{RR}{round-robin}
\newacronym{DA}{DA}{Dutch auction}
\newacronym{AV}{AV}{autonomous vehicle}
\newacronym{CWFL}{CWFL}{clustered WFL}
\newacronym{WFL}{WFL}{wireless federated learning}
\newacronym{RSMA}{RSMA}{rate splitting multiple access}
\newacronym{IoT}{IoT}{Internet-of-Things}
\newacronym{TDMA}{TDMA}{time-domain multiple access}
\newacronym{NOMA}{NOMA}{non-orthogonal multiple access}
\newacronym{ML}{ML}{machine learning}
\newacronym{MIMO}{MIMO}{multiple-input multiple-output}
\newacronym{CT}{CT}{compute-then-transmit}
\newacronym{FP}{FP}{fractional programming}
\newacronym{CF-mMIMO}{CF-mMIMO}{cell free massive MIMO}
\newacronym{iid}{i.i.d.}{independent and identically distributed}
\newacronym{AD}{AD}{autonomous driving}
\newacronym{DL}{DL}{downlink}
\newacronym{UL}{UL}{uplink}
\newacronym{IC}{IC}{interference cancellation}
\newacronym{SIC}{SIC}{successive interference cancellation}
\newacronym{BS}{BS}{base station}
\newacronym{TX}{TX}{transmit}
\newacronym{RX}{RX}{receive}
\newacronym{MU}{MU}{multi-user}
\newacronym{SISO}{SISO}{single-input single-output}
\newacronym{AWGN}{AWGN}{additive white Gaussian noise}
\newacronym{SINR}{SINR}{signal-to-interference-and-noise ratio}
\newacronym{FL}{FL}{federated learning}
\newacronym{CPU}{CPU}{central processing unit}
\newacronym{KNN}{KNN}{K-nearest-neighbor}
\newacronym{RF}{RF}{radio frequency}
\newacronym{GD}{GD}{gradient descent}
\newacronym{V2X}{V2X}{vehicle-to-anything}
\newacronym{i.i.d.}{i.i.d.}{independent and identically distributed}

\newacronym{RSS}{RSS}{received signal strength}
\newacronym{FIM}{FIM}{fisher information matrix}
\newacronym{ToA}{ToA}{time of arrival}
\newacronym{ToF}{ToF}{time of flight}
\newacronym{AoA}{AoA}{angle of arrival}
\newacronym{GP}{GP}{Gaussian process}
\newacronym{2D}{2D}{two-dimensional}
\newacronym{GPR}{GPR}{Gaussian process regression}
\newacronym{GNSS}{GNSS}{global navigation satellite systems}
\newacronym{B5G}{B5G}{beyond fifth-generation}
\newacronym{6G}{6G}{sixth-generation}
\newacronym{MP}{MP}{message-passing}
\newacronym{BP}{BP}{belief propagation}
\newacronym{EP}{EP}{expectation propagation}
\newacronym{GNN}{GNN}{graph neural network}
\newacronym{WLAN}{WLAN}{wireless local area network}

\newacronym{RRH}{RRH}{remote radio head}
\newacronym{GPS}{GPS}{Global Positioning System}
\newacronym{RFID}{RFID}{radio frequency identification}
\newacronym{TCAS}{TCAS}{traffic alert and collision avoidance systems}
\newacronym{RMSE}{RMSE}{root mean square error}
\newacronym{MSE}{MSE}{mean square error}

\newacronym{SGD}{SGD}{stochastic gradient descent}
\newacronym{PDF}{PDF}{probability density function}
\newacronym{CU}{CU}{computing unit}
\newacronym{DM-MIMO}{DM-MIMO}{distributed massive multiple-input multiple-output}
\newacronym{LOS}{LOS}{line-of-sight}
\newacronym{NLOS}{NLOS}{non-line-of-sight}
\newacronym{ROI}{ROI}{region of interest}
\newacronym{AP}{AP}{access point}
\newacronym{TDOA}{TDOA}{time difference of arrival}
\newacronym{UE}{UE}{user equipment}
\newacronym{dB}{dB}{decibel}
\newacronym{RIS}{RIS}{reconfigurable intelligent surface}
\newacronym{CG}{CG}{conjugate gradient}

\newacronym{PG}{PG}{proximal gradient}
\newacronym{SVT}{SVT}{singular value thresholding}
\newacronym{NN}{NN}{nuclear norm}
\newacronym{NMSE}{NMSE}{normalized mean square error}
\newacronym{MC}{MC}{matrix completion}
\newacronym{NP}{NP}{non-deterministic polynomial-time}
\newacronym{EDM}{EDM}{euclidean distance matrix}
\newacronym{SC}{SC}{soft-connected}
\newacronym{CRLB}{CRLB}{Cramér-Rao Lower Bound}
\newacronym{PoA}{PoA}{phase of arrival}
\newacronym{UAV}{UAV}{unmanned aerial vehicle}
\newacronym{VR}{VR}{virtual reality}
\newacronym{MDS}{MDS}{multidimensional scaling}
\newacronym{SMDS}{SMDS}{super multidimensional scaling}

\newacronym{RBL}{RBL}{rigid body localization}
\newacronym{RBT}{RBT}{rigid body tracking}
\newacronym{SC-RBL}{SC-RBL}{soft-connected RBL}
\newacronym{W-RBL}{W-RBL}{\underline{wireless} RBL}

\newacronym{SDP}{SDP}{semidefinite programming}
\newacronym{JCAS}{JCAS}{joint communication and sensing}
\newacronym{SDR}{SDR}{semi-definite relaxation}

\newacronym{OPP}{OPP}{orthogonal Procrustes problem}
\newacronym{SLAM}{SLAM}{simultaneous localization and mapping}
\newacronym{WLS}{WLS}{weighted least squares}
\newacronym{SI}{SI}{soft-impute}
\newacronym{GaBP}{GaBP}{Gaussian belief propagation}

\newacronym{SGA}{SGA}{scalar Gaussian approximation}

\newacronym{6D}{6D}{sixth-dimensional}
\newacronym{FDOA}{FDOA}{frequency difference of arrival}

\newacronym{wlg}{w.l.g.}{without loss of generality}
\newacronym{sIC}{soft-IC}{soft interference cancellation}
\newacronym{WL}{WL}{wireless localization}
\newacronym{XR}{XR}{extended reality}
\newacronym{IMU}{IMU}{inertial measurement unit}

\newacronym{LS}{LS}{least squares}

\let\ogfootnote\footnote
\RenewDocumentCommand{\footnote}{ O{8pt} O{-1pt} m }{\ogfootnote{%
   \setlength{\baselineskip}{#1}%
   \setlength{\lineskiplimit}{#2}%
#3}}

\begin{document}

\markboth{}{Niclas F\"uhrling {et al.}}

\makeatletter
\def\@maketitle{\newpage
  \bgroup%
    \vspace*{-10.5pt}%
    {\fontfamily{\sfdefault}\fontsize{7.5}{9}\selectfont\leftskip2pc plus1fill\rightskip2pc plus1fill
     {\bfseries\color{red}Please find the official IEEE \underline{published} version of this article on IEEE Xplore \href{https://ieeexplore.ieee.org/abstract/document/11680541}{\color{blue}[here]} and cite as:}\par
     {\color{NavyBlue}N.~F\"uhrling \emph{et al.}, ``Rigid Body Localization via Gaussian Belief Propagation With Quadratic Angle Approximation,'' \emph{IEEE Open Journal of Signal Processing}, vol.~7, pp.~944--956, Sept.~2026.}\par}%
     \vskip10pt{\titlefont{\textcolor{ojcolor}{\@title}}\par\vskip8.5pt}%
    {\authorfont\@author\par}\vskip4pt{\afffont\@affil}\par\@IEEEspecialpapernotice\vskip5pt%
    \ifx\@corresp\@empty\else{\corfont\@corresp\par}\vskip5.5pt\fi%
    \ifx\@authornote\@empty\else{\afffont\@authornote\par}\vskip27.5pt\fi%
    \centerline{\rule{36pc}{.5pt}}\par\vskip11.5pt%
    \ifvoid\abstractbox\else{\reset@font\box\abstractbox}\fi\par\vskip11.5pt%
    \ifvoid\keybox\else{\reset@font\box\keybox}\fi\par\vskip-32pt%
    \centerline{\rule{36pc}{.5pt}}
    \par\addvspace{33.5pt}\egroup%
    \let\maketitle\relax
}%
\makeatother

\title{Rigid Body Localization via\\ Gaussian Belief Propagation With\\ Quadratic Angle Approximation\vspace{-.5ex}}

\author{Niclas~F\"uhrling\authorrefmark{1}\textsuperscript{\orcidlink{0000-0003-1942-8691}}, Graduate Student Member, IEEE, \\ Hyeon Seok Rou\authorrefmark{1}\textsuperscript{\orcidlink{0000-0003-3483-7629}}, Member, IEEE, \\ Giuseppe Thadeu Freitas de Abreu\authorrefmark{1}\textsuperscript{\orcidlink{0000-0002-5018-8174}}, Senior Member, IEEE,\\ David~Gonz{\'a}lez~G.\authorrefmark{2}\textsuperscript{\orcidlink{0000-0003-2090-8481}}, Senior Member, IEEE, and Osvaldo~Gonsa\authorrefmark{2}\textsuperscript{\orcidlink{0000-0001-5452-8159}}}
\affil{School of Computer Science and Engineering, Constructor University, 28759 Bremen, Germany}
\affil{Wireless Communications Technologies Group, AUMOVIO Germany GmbH, 65936 Frankfurt am Main, Germany}
\corresp{Corresponding author: Niclas F\"uhrling (e-mail: nfuehrling@constructor.university).}

\begin{abstract}
\ac{GaBP} is a technique that relies on linearized error and input-output models to yield low-complexity solutions to complex estimation problems, which has been recently shown to be effective in the design of range-based \ac{GaBP} schemes for stationary and moving \ac{RBL} in \ac{3D} space, as long as {\color{black}the relative rotation between the prior position and the target rigid body is sufficiently small}.
In this article we present a novel range-based \ac{RBL} scheme via \ac{GaBP} that {\color{black}relaxes} the latter limitation {\color{black}significantly}.
To this end, the proposed method incorporates a quadratic angle approximation to linearize the relative orientation between the prior and the target rigid body, enabling high precision estimates of corresponding rotation angles even for large deviations.
Leveraging the resulting linearized model, we derive the corresponding \ac{MP} rules to obtain estimates of the translation vector and rotation matrix of the target rigid body, relative to a prior reference frame.
Numerical results corroborate the good performance of the proposed angle approximation itself, as well as the consequent \ac{RBL} performance in terms of \ac{RMSE} in comparison to the \ac{SotA}, while maintaining a low computational complexity.
\vspace{-3ex}
\end{abstract}

\begin{IEEEkeywords}
\vspace{1ex}
Rigid body localization, Gaussian belief propagation, small angle approximation.
\vspace{-2ex}
\end{IEEEkeywords}

\maketitle

\section{Introduction}

\IEEEPARstart{S}{ensing} \cite{Kong_2024} can be seen as one of the main applications in \ac{B5G} systems, as defined in the IMT-2030 \cite{02:00074} goals.
In particular, the emerging \ac{ISAC} paradigm of new wireless standards such as \ac{6G} and 802.11.bf \ac{WLAN}, demonstrate that wireless systems can use radio signals not only for communications, but also to perform sensing \cite{Zhang_2021,Rayan_Journal}.
A key sensing application is to localize and detect the shape and orientation of objects of interest in the environment, which can be performed via radio-based techniques \cite{Rou_TWC2024}.
Part of the larger sensing task is \ac{WL} \cite{Yassin_2016, GhodsTWC2018,Zafari_19}, whereby different properties of the radio signal such as finger-prints \cite{VoCST2016}, \ac{RSSI} \cite{Nic:RSSI}, \ac{AoA} \cite{Al-SadoonTAP2020}, delay \cite{ZengTSP2022} or combinations of those \cite{MacagnanoTWC2013} are used, to estimate the location of targets.
But more sophisticated sensing-driven applications such as autonomous driving and robotic systems require capabilities including navigation \cite{eckenhoff_2019}, collision detection \cite{Bruk_2023}, and vehicle trajectory and path prediction \cite{Huang_2022}, which in turn demand not only accurate position estimates of point targets, but more detailed and reliable information about the shape and orientation of \ac{3D} objects \cite{Yang_JMD09, Whittaker_IEEE06}.

This requirement gives rise to a localization paradigm known as \ac{RBL} \cite{EggertMVA1997, Diebel2006RigidBodyAttitude, Chepuri_2013, Chen_2015, Bras_2016, WangTSP2020, ZhaMRBL2021, Yu_ITJ23, fuehrling2025rigidbodylocalizationtracking, fuehrling2025robustegoisticrigidbody}. 
To elaborate, while traditional \ac{WL} schemes are designed to estimate the positions of individual point-targets independently, the objective of \ac{RBL} technology is to jointly estimate the global translation and attitude (jointly referred to as pose) of a \ac{3D} object, based on a collection of landmark points whose positions are fixed relative to one another due to the rigid shape of the object.
{\color{black} \ac{RBL}, therefore, is a solution to multiple of the aforementioned applications.
However, the most prominent application of \ac{RBL} in light of this work is tracking of rigid bodies, which has been studied extensively in the literature \cite{Chepuri_2013,eckenhoff_2019,Al-SadoonTAP2020}, where the position and orientation of a rigid body need to be estimated over time, such as in autonomous driving, where the accurate localization of vehicles is essential for safe navigation and collision avoidance.
While the relative deviation of the position and orientation from the previous position and orientation is typically small in tracking scenarios, there are cases where larger deviations can occur, such as during sudden maneuvers or in the presence of disturbances, which motivates the need for accurate and robust \ac{RBL} schemes.
}

Existing approaches for estimating the location and attitude of rigid objects span multiple sensing modalities.
For instance, computer vision-based methods typically rely on pose estimation from image or video data \cite{Chaoyi_ICASSP21, Xiang_arxiv17}. 
While effective in controlled environments, these approaches often require large data volumes and computationally intensive processing, which can limit their applicability in resource-constrained or real-time systems. 
Alternatively, \ac{IMU}-based techniques exploit measurements from accelerometers, gyroscopes, and magnetometers to infer orientation and motion \cite{AghiliTM2013, Zhao_JPCS18}. 
However, such methods are prone to drift, require frequent recalibration, and often depend on external aiding technologies, thereby partially undermining their self-contained nature.

In contrast to the aforementioned approaches, the wireless \ac{RBL} framework considered in this work \cite{EggertMVA1997, Diebel2006RigidBodyAttitude, Chen_2015, ZhaMRBL2021, Yu_ITJ23, fuehrling2025rigidbodylocalizationtracking} extends the conventional range-based \ac{WL} paradigm to the \ac{RBL} problem, by explicitly exploiting range measurements between landmark points\footnote{Landmark points can either represent edges and corners passively detected via radar-like methods, or actual sensors located at the rigid body. In this article we focus on \ac{RBL} algorithms rather than ranging methods, and such that both approaches can be abstracted as equivalent.} on the rigid body and a set of \textit{anchor} nodes at known positions \cite{AlcocerCDC2008, SandPWCS2014}. 
This formulation enables the joint estimation of the rigid body's global position, orientation, and, when applicable, its geometric configuration.
Earlier \ac{SotA} contributions to range-based \ac{RBL} approaches are based on either algebraic methods leveraging \ac{MDS} \cite{fuehrling2025robustegoisticrigidbody}, \ac{SMDS} \cite{fuehrling2025smdsbasedrigidbodylocalization} or \ac{SDR} \cite{Jiang2018, Wang2020}, all of which have a cubic computational complexity at least.

A representative example of a well-balanced solution that combines high estimation accuracy with moderate computational complexity is the \ac{LS} method proposed in \cite{Chen_2015}. 
In the scheme thereby, preliminary position estimates for the individual landmark points are first obtained from range measurements via a \ac{LS} formulation based on the linearized model in \cite{MaICASSP2011}. 
The rigid body translation and rotation are then extracted using a \ac{SVD}-based procedure, followed by a refinement step that enforces the rigid body constraints through an Euler-angle parametrization solved via \ac{WLS}.

The ability of the method in \cite{Chen_2015} to solve the \ac{RBL} problem within a unified framework stems from the use of linearized error models on the measured square distances \cite{MaICASSP2011, HoTSP2004}. 
Nevertheless, the reliance on \ac{LS} formulations and multi-stage processing can limit estimation performance, particularly in scenarios with strongly coupled variables \cite{Vaskevicius_LS2023}, as well as the fact that a small angle approximation is used in the linearized model. 
In terms of complexity, message-passing-based methods \cite{FengAMP2022} are well known to offer favorable accuracy-complexity trade-offs in such settings. 
These methods enable efficient inference either through alternating schemes \cite{ZhuAltSAMP2024} or through joint bilinear formulations \cite{ParkerBiGaMPI_2014, ParkerBiGaMPII_2014, Rou_TWC2024, Rou_Asilomar2022}.
There are multiple message passing-based localization techniques in the \ac{WL} literature, though not specifically designed for \ac{RBL}, for example, the method presented in \cite{Jin_2020}, where a \ac{RSSI}-based cooperative localization problem involving an unknown path loss exponent is solved via message passing algorithms that solve a probabilistic inference problem. 
In terms of sensing applications, \cite{Yuan_2016} employs \ac{GaBP} for simultaneous localization and synchronization based on \ac{ToA} measurements, while \cite{Yu_2021} leverages \ac{GaBP} for distributed \ac{MIMO} radar localization, resulting in low computational complexity and reduced communication cost.

The most recent, and to the best of the authors' knowledge, the only message passing-based methods specifically designed for \ac{RBL} methods are presented in \cite{Vizitiv_2025, fuehrling20246drigidbodylocalization}, where the linearized model in \cite{Chen_2015} is leveraged in the design of multiple linear and bilinear \ac{GaBP} algorithms for estimation of rigid body parameters both in stationary and moving scenarios, that is, yielding estimates for both the translation and rotation, as well as the angular and translational velocities of the rigid body.
However, a limitation of the linearized system model used in \cite{Chen_2015} and \cite{fuehrling20246drigidbodylocalization}, as mentioned above, is that it relies on the small angle approximation of the rotation matrix, which is only valid for very small angles, typically below $10^\circ$ \cite{Diebel2006RigidBodyAttitude} and significantly degrades the performance for larger angle deviations, as shown in \cite{fuehrling20246drigidbodylocalization}.

In view of the above, based on our work in \cite{Vizitiv_2025}, we propose in this article a \ac{GaBP}-based approach for \ac{RBL}, in which we use a novel quadratic approximation of the rotation matrix{\color{black}, retaining the most dominant second-order terms while preserving a form compatible with efficient \ac{GaBP} estimation,} instead of the commonly used small angle approximation, which is only valid for very small angles.
To that extent, we derive a linearized system model that uses the proposed quadratic angle approximation of the rotation matrix, which is then used in a modified bilinear \ac{GaBP} algorithm to estimate the rigid body parameters, $i.e.$, the translation and rotation of the target rigid body.

Our contributions can be summarized as follows \begin{itemize}
  {\color{black}\item A novel vectorized {\color{black}approximation} of the \ac{3D} rotation matrix via a quadratic angle approximation is proposed, which is valid for {\color{black}larger} angle deviations between the orientations of the prior and true target{\color{black}, compared to the small angle approximation applied in the \ac{SotA}}.
  \item A new linearized system model for \ac{RBL} is derived based on the proposed {\color{black}approximated vectorized rotation matrix} that relates the range measurements {\color{black}of each landmark point} directly to the rigid body parameters and enables the use of \ac{GaBP}-based estimation.}
  \item A new bilinear \ac{GaBP} algorithm is presented to estimate the rigid body parameters, $i.e.$, translation and rotation, based on the proposed linearized system model.
\end{itemize}
\vspace{-2ex}

The structure of the remainder of the article is as follows.
First, the rigid body system model, as well as the measurement model is described in Section \ref{sec:prior}.
Next, in Section \ref{sec:prop}, the proposed linearized system model via the quadratic angle approximation is introduced, followed by the modified \ac{GaBP} algorithm used to estimate the rigid body parameters.
Finally, a comparison of the proposed scheme with the small angle approximation, \ac{GaBP}-based \ac{RBL} method is presented in Section \ref{sec:res}.

\section{Rigid Body Localization System Model}
\label{sec:prior}

\subsection{Rigid Body System Model}

\begin{figure}[H]
\vspace{-2ex}
\centering
\includegraphics[width=1\columnwidth]{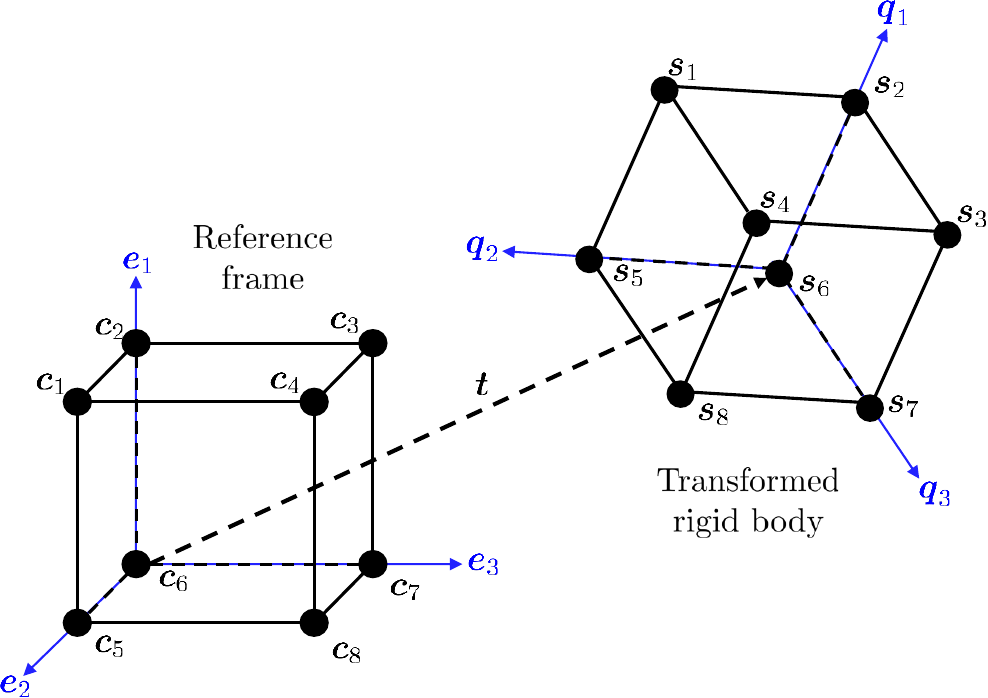}    
\vspace{-3ex}
\caption[]{An illustration of the transformation of a rigid body, whose location has a \ac{3D} rotation $\boldsymbol{Q}$ and a translation $\boldsymbol{t}$ applied to the reference frame, as determined by \eqref{eq:basic_model_RB}.}
\label{fig:RB_trans_plot}
\end{figure}

Consider, as illustrated in Fig.~\ref{fig:RB_trans_plot}, a scenario where a rigid body consisting of $N$ landmark points is transformed -- that is changes its pose -- in \ac{3D} Euclidean space.
Each sensor of the rigid body is described by a $3 \times 1$ vector denoted by $\boldsymbol{c}_n \in \mathbb{R}^{3\times 1}$ for $n=\{1, \ldots, N\}$.
The conformation matrix, describing the initial reference position of the rigid body is therefore defined by $\boldsymbol{C}=\left[\boldsymbol{c}_{1}, \boldsymbol{c}_{2}, \ldots, \boldsymbol{c}_{N}\right] \in \mathbb{R}^{3 \times N}$ at the reference frame (local axis) of the rigid body.

To define the transformation of the rigid body in \ac{3D} space, a translation and rotation needs to be introduced, respectively described by the translation vector $\boldsymbol{t} \triangleq [t_x, t_y, t_z]^\intercal \in\mathbb{R}^{3\times 1}$ consisting of the translation distances for each axis, and a \ac{3D} rotation matrix\footnote{The rotation matrix $\boldsymbol{Q}$ is part of the special orthogonal group such that $SO(3)=\left\{\boldsymbol{Q} \in \mathbb{R}^{3 \times 3}: \boldsymbol{Q}^\intercal \boldsymbol{Q} = \mathbf{I}, ~\mathrm{det}(\boldsymbol{Q})=1\right\}$ \cite{Diebel2006RigidBodyAttitude} needs to hold.} $\boldsymbol{Q} \in \mathbb{R}^{3\times 3}$ given by \eqref{eq:rotation_matrix}, where $\bm{Q}_{x}, \,\bm{Q}_{y}, \,\bm{Q}_{z} \in \mathbb{R}^{3\times 3}$ are the roll, pitch, and yaw rotation matrices about the \textit{x-,y-,z-}axes by rotation angles of $\theta_x, \theta_y, \theta_z \in [-180^\circ, 180^\circ]$ degrees, respectively{\color{black}, which, however, are limited to smaller angles in the approximation, such that $\theta_x, \theta_y, \theta_z \in [-\pi/4, \pi/4]$ radians.}

\begin{figure*}[t!]
\setcounter{equation}{0}
\normalsize
\begin{eqnarray}
\label{eq:rotation_matrix}
\bm{Q} \triangleq \bm{Q}_{z}(\theta_z)\,\bm{Q}_{y}(\theta_y)\,\bm{Q}_{x}(\theta_x)
=\!\!\!\text{\scalebox{1}{$\overbrace{\left[
\begin{array}{@{}c@{\;\,}c@{\;\,}c@{}}
\cos\theta_z&-\sin\theta_z& 0\\
\sin\theta_z& \cos\theta_z& 0\\
0 	    & 0           & 1\\
\end{array}\right]}^{\triangleq \,\bm{Q}_z \in\, \mathbb{R}^{3\times 3}}\!\cdot\!
\overbrace{\left[
\begin{array}{@{}c@{\;\,}c@{\;\,}c@{}}
\cos\theta_y & 0           & \sin\theta_y\\
0			& 1			  & 0\\
-\sin\theta_y& 0 		  & \cos\theta_y\\
\end{array}\right]}^{\triangleq \,\bm{Q}_y \in\, \mathbb{R}^{3\times 3}}\!\cdot\!
\overbrace{\left[
\begin{array}{@{\,}c@{\;\,}c@{\;\,}c@{\!}}
1 			& 0			  & 0\\
0			& \cos\theta_x& -\sin\theta_x\\
0			& \sin\theta_x& \cos\theta_x\\
\end{array}\right]}^{\triangleq \bm{Q}_x \in\, \mathbb{R}^{3\times 3}}$}}\\
&&\hspace{-87ex}
=\!\!\text{\scalebox{1}{$
\left[
\begin{array}{@{\,}c@{\;\,}c@{\;\,}c@{\,}}
\cos\theta_y\cos\theta_z & \sin\theta_x\sin\theta_y\cos\theta_z- \cos\theta_x\sin\theta_z & \cos\theta_x\sin\theta_y\cos\theta_z+\sin\theta_x\sin\theta_z\\
\cos\theta_y\sin\theta_z & \sin\theta_x\sin\theta_y\sin\theta_z+ \cos\theta_x\cos\theta_z & \cos\theta_x\sin\theta_y\sin\theta_z-\sin\theta_x\cos\theta_z\\
-\sin\theta_y			& \sin\theta_x\cos\theta_y								  & \cos\theta_x\cos\theta_y\\
\end{array}\right]$}}
=\!\!\left[
\begin{array}{@{\,}c@{\;\,}c@{\;\,}c@{\,}}
q_{1,1} & q_{1,2} & q_{1,3}\\
q_{2,1} & q_{2,2} & q_{2,3}\\
q_{3,1} & q_{3,2} & q_{3,3}\\
\end{array}\right]\!\!.\nonumber
\end{eqnarray}
\setcounter{equation}{1}
\hrulefill
\end{figure*}

\begin{figure}[H]
\centering
\includegraphics[width=0.8\columnwidth]{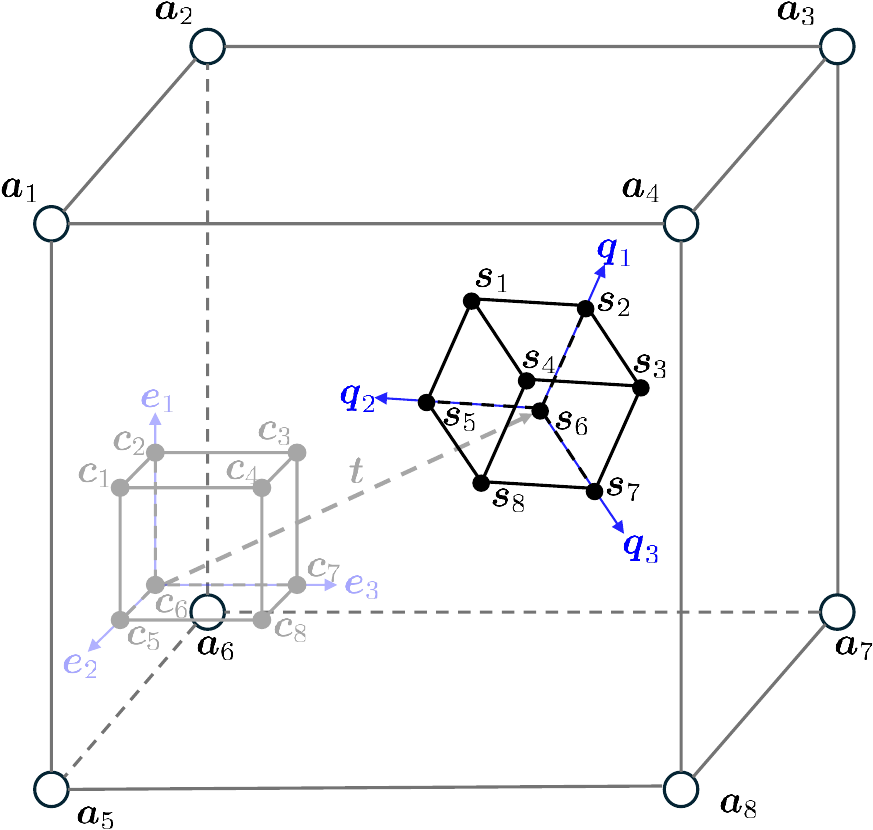}    
\caption[]{Illustration of the transformed rigid body surrounded by $M$ anchors, whose locations are known.}
\label{fig:RB_anchor}
\end{figure}

Next, the transformed coordinates of the $n$-th sensor after the rotation and translation is described by
\begin{equation}
\label{eq:basic_model_RB}
\boldsymbol{s}_{n} =\boldsymbol{Q} \boldsymbol{c}_{n}+\boldsymbol{t} \in \mathbb{R}^{3 \times 1},
\end{equation}
which is applied identically to all $N$ landmark points of the rigid body, as illustrated in Fig.~\ref{fig:RB_trans_plot}, such that for the full body, the equation can be rewritten as
\begin{equation}
\label{eq:basic_model_RB_Stack}
\boldsymbol{S} =\boldsymbol{Q} \boldsymbol{C}+\boldsymbol{t}\bm{1}_N^\intercal \in \mathbb{R}^{3 \times N},
\end{equation}
where $\bm{1}_N$ denotes a column vector of length $N$.

Finally, as illustrated in Fig.~\ref{fig:RB_anchor}, in order to perform measurements for the localization, the target rigid body is surrounded by a total of $M$ reference sensors (hereafter referred to  simply as anchors), defined by their known location $\boldsymbol{a}_m \in \mathbb{R}^{3\times 1}$ for $m=\{1, \ldots, M\}$.

\subsection{Measurement Model}

The pairwise range measurements between anchors and landmark points considered for \ac{RBL}, assumed to be available by a common processing unit, {\color{black}which collects the measurements of all $M$ anchors and $N$ landmark points} are described by
\begin{equation}
\label{eq:range_model}
\tilde{d}_{m,n} = d_{m,n} + w_{m,n} = \left\|\boldsymbol{a}_{m} - \boldsymbol{s}_{n}\right\|_{2} + w_{m,n} \in \mathbb{R},
\end{equation}
where corresponding conventional squared range measurements are modeled as
\begin{equation}
\label{eq:sq_range_meas}
\tilde{d}_{m,n}^2 = \left\|\boldsymbol{a}_{m} - \boldsymbol{s}_{n}\right\|_{2}^2 + 2d_{m,n}w_{m,n} + w_{m,n}^2  \in \mathbb{R},
\end{equation}
with $w_{m,n} \sim \mathcal{N}(0, \sigma_w^2)$ denoting the \ac{i.i.d.} \ac{AWGN} of  variance $\sigma_w^2$ affecting the range measurement, while $d_{m,n} \triangleq \left\|\boldsymbol{a}_{m} - \boldsymbol{s}_{n}\right\|_{2}$ is the true Euclidean distance between the $m$-th anchor and the $n$-th sensor.

Following \cite{Chen_2015,HoTSP2004,MaICASSP2011}, the range measurement equation in \eqref{eq:sq_range_meas} can be reformulated in terms of a linear relation with a \textit{composite ranging noise} $\xi_{n} \in \mathbb{R}$ defined as
\begin{equation}
\label{eq:pos_lin_eq}
\xi_{m,n} = \tilde{d}_{m,n}^{2} - \left\|\boldsymbol{a}_{m}\right\|^{2}_2 -\left\|\boldsymbol{s}_{n}\right\|^{2}_2 +2 \boldsymbol{a}_{m}^\intercal \boldsymbol{s}_{n} \approx 2d_{m,n}w_{m,n},
\end{equation}
where the second-order noise term $w_{m,n}^2$ is neglected.{\color{black}\footnote{\color{black}The second-order noise term, often neglected in linear approximations, can introduce non-negligible biases in nonlinear estimation problems{\color{black}, here amounting to a fixed offset $\mathbb{E}[w_{m,n}^2]=\sigma_w^2$}. Its influence is typically minor in large scale systems such as the scenario considered in this work, and may become significant primarily in small scale systems involving nonlinear estimation under high noise conditions.}}

{\color{black}The resulting composite noise $\xi_{m,n} \approx 2 d_{m,n} w_{m,n}$ is distance dependent{\color{black}, $i.e.$, heteroscedastic across anchor-landmark pairs,} with variance $4 d_{m,n}^2 \sigma_w^2$. In the message passing updates, this distance dependent variance is approximated by an effective constant variance $N_0$, following the standard linearized treatment of \cite{Chen_2015,HoTSP2004}, which preserves the closed form Gaussian belief propagation recursions{\color{black}, and whose value used in the simulations is evaluated for the realized noise draw for that specific Monte Carlo trial}.}

By rearranging and stacking \eqref{eq:pos_lin_eq} for all $M$ anchors, the system can be reformulated as a linear system on the $n$-th unknown sensor variable, which yields
\vspace{1ex}
\begin{align}
\boldsymbol{y}_{n} \!&\triangleq\!\!\!
\begin{bmatrix}
\!\tilde{d}_{1,n}^{2} \!\!-\! \left\|\boldsymbol{a}_{1}\right\|^{2}_2 \!\\
\!\vdots \\
\tilde{d}_{M,n}^{2} \!\!-\! \left\|\boldsymbol{a}_{M}\right\|^{2}_2 \\
\end{bmatrix} 
\!\!\!=\!\!\!
\underbrace{
\begin{bmatrix}
\!-2 \boldsymbol{a}_{1}^\intercal, &\!\!\!\! 1\; \\
\!\vdots &\!\!\!\! \vdots~\\
\!-2 \boldsymbol{a}_{M}^\intercal, &\!\!\!\! 1\; \\
\end{bmatrix}}_{\triangleq \,\boldsymbol{G} \,\in\, \mathbb{R}^{M \!\times\!4}}
\!\!\!\!\!\!\!\overbrace{\!
\begin{bmatrix}
\boldsymbol{s}_n \\[1ex]
||\boldsymbol{s}_n||_2^2
\end{bmatrix}\!}^{ ~~~\triangleq \, \boldsymbol{x}_n \, \in \, \mathbb{R}^{4 \!\times\!1}}
\!\!\!\!\!\!+\!\!\!\!\!\!
\underbrace{\!
\begin{bmatrix}
\!\xi_{1,n}\!\\
\!\vdots\!\\
\!\xi_{M,n}\!\\
\end{bmatrix}
\!}_{\triangleq \,\boldsymbol{\xi}_n \,\in\, \mathbb{R}^{M \!\times\!1}}\!\!\!\!\!\!\! \in \!\mathbb{R}^{M \times 1}\!, \nonumber \\[-4ex] 
\label{eq:linear_sys}
\vspace{1ex}
\end{align}
where, $\boldsymbol{y}_{n} \in \mathbb{R}^{M \times 1}$ denotes the observed data vector, $\boldsymbol{G}$ $\in \mathbb{R}^{M \times 4}$ denotes the effective channel matrix, which has been implicitly defined to be constructed from range measurements and anchor positions $\bm{a}_m$.

In the above, the vector $\boldsymbol{x}_n \in \mathbb{R}^{4 \times 1}$ is implicitly defined, which contains the unknown coordinates of the $n$-sensor, as well as its distance to the origin, while the vector $\boldsymbol{\xi}_{n} \in \mathbb{R}^{M \times 1}$ gathers the composite noise quantities defined in \eqref{eq:pos_lin_eq}.

As shown in \cite{fuehrling20246drigidbodylocalization}, {\color{black} as a preliminary step,} the linear system in \eqref{eq:linear_sys} can be leveraged for the estimation of the unknown sensor coordinate vector $\boldsymbol{s}_n$ and sensor position norm $||\boldsymbol{s}||_2^2$ in $\boldsymbol{x}_n$ via a linear \ac{GaBP} approach{\color{black}, which is later required as an initialization for the subsequent rigid body parameter estimation}.
With the estimate of the sensor positions in hand, \eqref{eq:basic_model_RB} can either be invoked for the translation and rotation extraction via Procrustes analysis or other classical algorithms \cite{EggertMVA1997}, as described in \cite{Chen_2015}, or a bivariate \ac{GaBP} approach can be invoked to estimate the rigid body parameters directly, as described in \cite{Vizitiv_2025}, where a small angle approximation was used to linearize the rotation matrix to enable the \ac{GaBP} approach{\color{black}, a bivariate structure that the proposed method in Section \ref{sec:prop} deliberately retains,  to enable a controlled comparison between the two angle approximations}.
\subsection{SotA Parameter-based System Model via Small Angle Approximation}
\label{sec:reformulation}
\vspace{1ex}

As proposed in \cite{fuehrling20246drigidbodylocalization}, \eqref{eq:linear_sys} can be reformulated, such that system variables can be expressed directly in terms of the \ac{RBL} transformation parameters, $i.e.$, the \ac{3D} rotation angles $\boldsymbol{\theta} \triangleq [\theta_x, \theta_y, \theta_z]^\intercal \in \mathbb{R}^{3 \times 1}$ and translation vector $\boldsymbol{t}$ \cite{ZhaMRBL2021}.
To enable the reformulation, first, a small-angle approximation\footnotemark \cite{Diebel2006RigidBodyAttitude} is applied onto the rotation matrix of \eqref{eq:rotation_matrix}, obtained by the approximations $\cos\theta \approx 1$ and $\sin\theta \approx \theta$, which yields
\vspace{1ex}
\begin{eqnarray}
\label{eq:q_small_angle}
~~~~\bm{Q} \approx\!\!\left[\begin{array}{ccc}
1 & \theta_z & -\theta_y \\
-\theta_z & 1 & \theta_x \\
\theta_y & -\theta_x & 1
\end{array}\right] \in \mathbb{R}^{3 \times 3},
\vspace{1ex}
\end{eqnarray}
which can be vectorized into a linear system directly in terms of the Euler angles $\bm{\theta}$ \cite{Chen_2015}, given by
\vspace{1ex}
\begin{eqnarray}
\mathrm{vec}(\boldsymbol{Q}) = \boldsymbol{\gamma} + \boldsymbol{L} \boldsymbol{\theta} = \overbrace{
\begin{bmatrix}
1 & 0 & 0 & 0 & 1 & 0 & 0 & 0 & 1
\end{bmatrix}^\intercal}^{\triangleq\, \boldsymbol{\gamma} \, \in \, \mathbb{R}^{9 \times 1}}&& \nonumber \\
&&
\label{eq:q_vec}
\hspace{-48ex} + \underbrace{
\begin{bmatrix}
0 &  1 & 0 & -1 & 0 & 0 & 0 & 0 & 0 \\ 
0 &  0 & -1 & 0 & 0 & 0 & 1 & 0 & 0 \\ 
0 &  0 & 0 & 0 & 0 & 1 & 0 & -1 & 0 \\ 
\end{bmatrix}^\intercal\!\!\!}_{\triangleq\, \boldsymbol{L} \, \in \, \mathbb{R}^{9 \times 3}} 
\cdot\!
\begin{bmatrix}
\theta_x \\ \theta_y \\ \theta_z
\end{bmatrix}\!\!.
\vspace{1ex}
\end{eqnarray}

\footnotetext{For practical rigid body tracking applications, subsequent transformation estimations can be assumed to be performed within a sufficiently short time interval such that the change in rotation angle remains small. Although the approximation remains valid for rotation angles up to approximately twenty degrees, convergence of the algorithm can still be achieved for larger angles. However, in such cases, the estimation accuracy may degrade.}

Next, when \eqref{eq:q_vec} is substituted into \eqref{eq:basic_model_RB} and \eqref{eq:pos_lin_eq} and the terms are rearranged, an alternate representation of the composite noise is obtained, given by
%
%
\vspace{1ex}
\begin{eqnarray}
\xi_{n} \!=&& \hspace{-4ex} \tilde{d}_{m,n}^{2}\!-\!\left\|\boldsymbol{a}_{m}\right\|^{2}_2\! -\! \left\|\boldsymbol{s}_{n}\right\|^{2}_2 \!+\! 2\!\left[\boldsymbol{c}_{n}^\intercal \otimes \boldsymbol{a}_{m}^\intercal \right]\!\boldsymbol{\gamma}\!\nonumber \\
&&\hspace{-4ex}+\!2\!\left[\boldsymbol{c}_{n}^\intercal \otimes \boldsymbol{a}_{m}^\intercal\right]\!\boldsymbol{L}\boldsymbol{\theta} +2 \boldsymbol{a}_{m}^\intercal \boldsymbol{t} \in \mathbb{R},\!\! \label{eq:delta_lin_eq}
\vspace{1ex}
\end{eqnarray}
where $\otimes$ denotes the Kronecker product operator and vectorization identity of the matrix product $\mathrm{vec}(\mathbf{X Y Z}) = (\mathbf{Z}^\intercal \otimes \mathbf{X}) \mathrm{vec}(\mathbf{Y})$ has been used.

\newpage
Finally, the fundamental system can be rewritten as a linearization of \eqref{eq:delta_lin_eq} in terms of the rigid body parameters $\boldsymbol{\theta}$ and $\boldsymbol{t}$, which yields
\begin{subequations}
\label{eq:delta_t_lin_syst}
\begin{equation}
\boldsymbol{z}_{n} = \boldsymbol{H}_{\theta} \!\cdot\! \boldsymbol{\theta} + \boldsymbol{H}_{t} \!\cdot\! \boldsymbol{t} + \boldsymbol{\xi}_{n} \in \mathbb{R}^{M \times 1},
\end{equation}
with
\begin{equation}
\boldsymbol{z}_{n} \!=\!\! \left[\!\!\begin{array}{c}
\tilde{d}_{1,n}^{2}-\left\|\boldsymbol{a}_{1}\right\|^{2}_2- \left\|\boldsymbol{s}_{n}\right\|^{2}_2 + 2\left[\boldsymbol{c}_{n}^\intercal \otimes \boldsymbol{a}_{1}^\intercal \right]\!\boldsymbol{\gamma} \\[1ex]
\vdots \\[1ex]
\tilde{d}_{M,n}^{2}-\left\|\boldsymbol{a}_{M}\right\|^{2}_2 - \left\|\boldsymbol{s}_{n}\right\|^{2}_2 + 2\left[\boldsymbol{c}_{n}^\intercal \otimes \boldsymbol{a}_{M}^\intercal \right]\!\boldsymbol{\gamma}
\end{array}\!\!\right] \!\! \in \mathbb{R}^{M \times 1},
\end{equation}
where $\boldsymbol{z}_{n} \in \mathbb{R}^{M \times 1}$ denotes the effective observed data vector, $\boldsymbol{\xi}_{n} \in \mathbb{R}^{M \times 1}$ is the vector of composite noise variables from \eqref{eq:pos_lin_eq}, and $\boldsymbol{H}_{\theta}$ and $\boldsymbol{H}_{t}$ are the two effective channel matrices for rotation and translation respectively, given by
\vspace{-3ex}
\begin{figure}[H]
\centering
{{\includegraphics[width=\columnwidth]{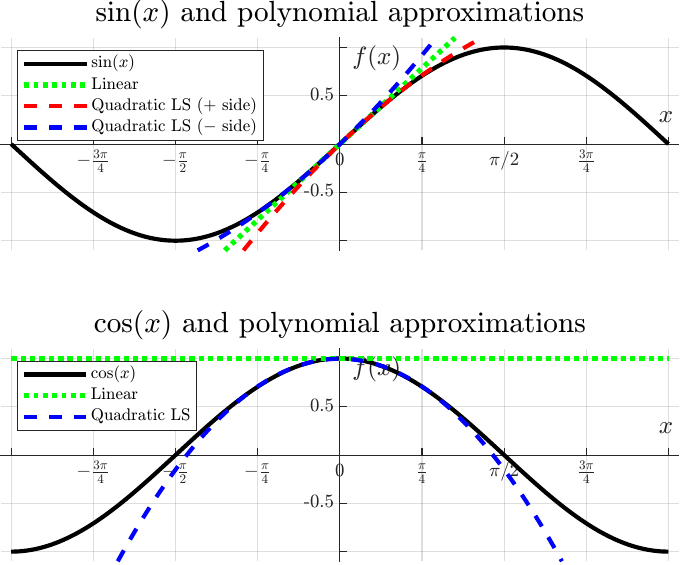}}}
\vspace{-3ex}
\caption[]{Various Sine and Cosine approximations.}
\label{fig:Approx}
\vspace{3ex}
{{\includegraphics[width=\columnwidth]{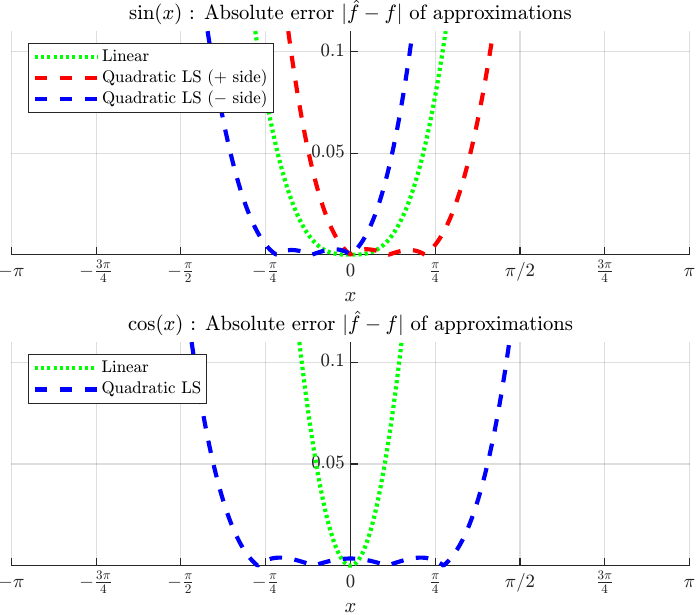}}}
\vspace{-3ex}
\caption[]{Errors of the Sine and Cosine approximations.}
\label{fig:ApproxERR}
\end{figure}
\begin{equation}
\boldsymbol{H}_{\theta} \!=\!\! \left[\begin{array}{c}
\!\!\!\!\!-2\!\left[{\color{black}\boldsymbol{c}_{n}}^\intercal \otimes \boldsymbol{a}_{1}^\intercal \right]\!\boldsymbol{L}\!\!\!\\
\vdots   \\[0.5ex]
\!\!\!\!-2\!\left[{\color{black}\boldsymbol{c}_{n}}^\intercal \otimes \boldsymbol{a}_{M}^\intercal \right]\!\boldsymbol{L}\!\!\!
\end{array}\right] \!\! \in \!\mathbb{R}^{M \times 3}\!,
\end{equation}
and 
\begin{equation}
  \vspace{1ex}
~ \boldsymbol{H}_{t} \!=\!\! \left[\begin{array}{c}
\!\!\!\!-2 \boldsymbol{a}_{1}^\intercal\!\!\!\!  \\
\vdots \\
\!\!\!\!-2 \boldsymbol{a}_{M}^\intercal\!\!\! 
\end{array}\right]\!\! \in \!\mathbb{R}^{M \times 3}.
\end{equation}
\end{subequations}

\vspace{-3ex}
\section{Proposed Method}
\label{sec:prop}

In light of the above, the proposed method proposes a reformulation of the linear system proposed in \cite{fuehrling20246drigidbodylocalization} by the introduction of a quadratic angle approximation, which is then used in a modified \ac{GaBP} algorithm to estimate the rigid body parameters directly.

\vspace{-2ex}
\subsection{System Model via Quadratic Angle Approximation}

Instead of the conventional small angle approximation, where $\sin\theta \approx \theta$ and $\cos\theta \approx 1$, we propose a novel quadratic approximation of the rotation matrix.\footnote{\color{black}Note that while the approximated rotation matrix does not satisfy the SO(3) constraints, the individual yaw, pitch, and roll angles can be used to reconstruct a valid rotation matrix after the final estimates are obtained.}
To that extent, we approximate the elements of a rotation matrix $\mathbf{Q}^{[i]} \in \mathbb{R}^{3 \times 3}$ using a quadratic approximation, based on

\begin{subequations}
  \begin{equation}
    \sin(\theta) \approx \alpha\, \theta^{[i-1]} \theta^{[i]} + \beta\, \theta^{[i]},
  \end{equation}
  \begin{equation}
    \cos(\theta) \approx \gamma - \delta\, \theta^{[i-1]} \theta^{[i]},
  \end{equation}
\end{subequations}
where the values for the parameters $\alpha, \beta, \gamma, \delta \in \mathbb{R}$ are chosen {\color{black} via a least-squares fit} such that the approximation is as accurate as possible in the interval $\theta \in [-\pi/4, \pi/4]$.
To that extent, for the sine approximation, we use $\alpha^+$ for positive angles and $\alpha^-$ for negative angles, while for the cosine approximation the same parameters are used for positive and negative angles.
In order to ensure that the approximations pass through the origin and the exact value at $\pi/4$ {\color{black} rad}, as shown in Fig.~\ref{fig:Approx}, the parameters are chosen as described in Table \ref{tab:approx_params}{\color{black}, via a conventional least squares approach}.
Finally, in order to linearize the model to be applicable in the proposed \ac{GaBP}, however, \(\theta^{[i-1]}\) has to be known from the previous iteration and \(\theta^{[i]}\) is the current unknown.

Following the procedure as in \eqref{eq:q_vec}, as derived in Appendix \ref{app:Derivation}, we then represent the vectorized matrix as an affine-linear function of $\boldsymbol{\theta}^{[i]}$, which yields
\begin{equation}
\operatorname{vec}\bigl(\mathbf{Q}^{[i]}\bigr)
=
\boldsymbol{\gamma}
+
\Big(\beta\,\mathbf L
      +\alpha\,\mathbf L\boldsymbol\Theta
      -\delta\,\mathbf D\boldsymbol\Theta \Big)\boldsymbol{\theta}^{[i]},
\label{eq:quad_approx}
\end{equation}
\begin{table}[H]
\centering
\caption[]{Parameters used for the sine and cosine approximations.}
\label{tab:approx_params}
\begin{tabular}{|l|c|c|}
\hline
Parameter & Value & Used for \\[0.5ex]
\hline\hline
$\alpha^+$ & $-0.16884$ & sine ($\theta > 0$) \\
\hline
$\alpha^-$ & $+0.16884$ & sine ($\theta < 0$) \\
\hline
$\beta$    & $1.03912$  & {\color{black}sine} \\
\hline
$\gamma$   & $\tfrac{577}{579}$ & cosine \\
\hline
$\delta$   & $\tfrac{274}{589}$ & cosine \\
\hline
\end{tabular}
\end{table}
where $\boldsymbol{\theta}^{[i]} = [\theta_x^{[i]}, \theta_y^{[i]}, \theta_z^{[i]}]^\intercal \in \mathbb{R}^{3\times1}$ denotes the unknown current angles and $\boldsymbol{\theta}^{[i-1]} = [\theta_x^{[i-1]}, \theta_y^{[i-1]}, \theta_z^{[i-1]} ]^\intercal \in \mathbb{R}^{3\times1}$ the known previous angles. 
The diagonal matrix $\boldsymbol{\Theta} = \operatorname{diag}(\boldsymbol{\theta}^{[i-1]}) \in \mathbb{R}^{3 \times 3}$ collects the previous angles on its diagonal. 
Furthermore, the constant vector is given by $\boldsymbol{\gamma} = \gamma^2 \cdot \operatorname{vec}(\mathbf{I}_3) = \gamma^2 [1, 0,  0,  0,  1,  0,  0,  0,  1 ]^\intercal$. 
The matrices $\mathbf{L}, \mathbf{D} \in \mathbb{R}^{9 \times 3}$ are defined as

\begin{subequations}
\begin{equation}
  \mathbf{L} =
\begin{bmatrix}
0 & 0 & 0 & 0 & 0 & \gamma & 0 & -\gamma & 0 \\
0 & 0 & -1 & 0 & 0 & 0 & 1 & 0 & 0 \\
0 & \gamma & 0 & -\gamma & 0 & 0 & 0 & 0 & 0
\end{bmatrix}^{\intercal},
\end{equation}
\begin{equation}
  \mathbf{D} =
\begin{bmatrix}
0 & 0 & 0 & 0 & 1 & 0 & 0 & 0 & 1 \\
1 & 0 & 0 & 0 & 0 & 0 & 0 & 0 & 1 \\
1 & 0 & 0 & 0 & 1 & 0 & 0 & 0 & 0
\end{bmatrix}^{\intercal}.
\end{equation}
\end{subequations}

As shown in Fig.~\ref{fig:ApproxERR}, the quadratic approximation yields a significantly lower error compared to the small angle approximation, especially for angles larger than $20^\circ$.
In particular, for the cosine approximation, the maximum error for angles up to $45^\circ$ is reduced from approximately $0.08$ to less than $0.005$.
For the sine approximation, the maximum error is reduced from approximately $0.3$ to less than $0.004$.

{\color{black}For the sake of discussion, we define the valid operating range as the largest orientation deviation for which the absolute error remains below $5 \times 10^{-3}$, the proposed approximation extends this range from approximately $18^\circ$ to $45^\circ$ for the sine and from $6^\circ$ to $54^\circ$ for the cosine, corresponding to an improvement of roughly $2.5\times$ and $9\times$, respectively.
While the approximation error is still increasing for angles larger than $45^\circ$, the proposed approximation degrades slowly and continues to yield a significantly lower error compared to the small angle approximation, with the sine and cosine errors at $60^\circ$ remaining approximately $5\times$ and $37\times$ smaller, respectively.
However, the corresponding errors could be reduced further by using higher order approximations, which is left for future work.}


\vspace{-2ex}
\subsection{Proposed Linear Model}
In light of the above, the fundamental system in \eqref{eq:delta_t_lin_syst} can be rewritten leveraging the linear quadratic approximation of \eqref{eq:quad_approx}, which yields
\begin{subequations}
\label{eq:delta_t_lin_syst_NEW}
\begin{equation}
\boldsymbol{z}_{n} = \boldsymbol{H}_{\theta} \!\cdot\! \boldsymbol{\theta} + \boldsymbol{H}_{t} \!\cdot\! \boldsymbol{t} + \boldsymbol{\xi}_{n} \in \mathbb{R}^{M \times 1},
\end{equation}
where $\boldsymbol{z}_{n} \in \mathbb{R}^{M \times 1}$, $\boldsymbol{\xi}_{n} \in \mathbb{R}^{M \times 1}$, and $\boldsymbol{H}_{t}$, are identical to the previous terms, while the effective channel matrix for the rotation is redefined as

\begin{equation}
\boldsymbol{H}_{\theta} \!=\!\! \left[\begin{array}{c}
\!\!\!\!\!-2\!\left[{\color{black}\boldsymbol{c}_{n}}^\intercal \otimes \boldsymbol{a}_{1}^\intercal \right]\!\bigl(\beta\,\mathbf L
      +\alpha\,\mathbf L\boldsymbol\Theta
      -\delta\,\mathbf D\boldsymbol\Theta\bigr)\!\!\!\\
\vdots   \\[0.5ex]
\!\!\!\!-2\!\left[{\color{black}\boldsymbol{c}_{n}}^\intercal \otimes \boldsymbol{a}_{M}^\intercal \right]\!\bigl(\beta\,\mathbf L
      +\alpha\,\mathbf L\boldsymbol\Theta
      -\delta\,\mathbf D\boldsymbol\Theta\bigr)\!\!\!
\end{array}\right] \!\! \in \!\mathbb{R}^{M \times 3}\!.
\end{equation}
\label{H_new}
\end{subequations}

Finally, it has to be ensured that, depending on the prior angles $\bm{\theta}^{[i-1]}$, the correct approximation parameters is used, as shown by Fig.~\ref{fig:Approx}, where there exists a plus and minus side for the approximation.


\vspace{-2ex}
\subsection{Bivariate GaBP for Rigid Body Parameter Estimation}
\label{sec:del_t_est_sol}
With the linear formulation of the system model in hand, the message passing rules for the \ac{GaBP} iterations need to be defined, which are similar to the algorithm shown in \cite{fuehrling20246drigidbodylocalization}, where in the case of transformation parameter estimation using the new model based on \eqref{eq:delta_t_lin_syst_NEW}, there exist two sets of variables $\theta_k$ with $k \in \{1,\ldots,K\}$ and $t_\ell$ with $\ell \in \{1,\ldots,K\}$, where $K$ is the dimension of the rigid body parameters to be estimated, $i.e.$, $K=3$ in \ac{3D} space, such that the \ac{GaBP} rules are elaborated separately, updating the model every iteration.
{\color{black}Furthermore, for the sake of convenience, the subscript $n$, which denotes the index of the $n$-th sensor node, shall temporarily be dropped, as the derivation is identical for all $N$ sensor nodes.}

To perform the \ac{GaBP} estimation, first \ac{sIC} is performed on the observed information for rigid body parameters, $i.e.$, the rotation angles and translation as
\begin{subequations}
\label{eq:del_t_soft_IC}
\begin{align}\label{eq:del_IC}
\tilde{z}_{\theta:m,k}^{[j]} &= z_{m} - \sum_{i \neq k} h_{\theta:m,i}\hat{\theta}_{m,i}^{[j]} - \sum_{i = 1}^{K} h_{t:m,i}\hat{t}_{m,i}^{[j]} \\[-1ex]
&= h_{\theta:m,k}\theta_{k} +  \sum_{i = 1}^{K} h_{t:m,i}(t_{i} - \hat{t}_{m,i}^{[j]}) \nonumber \\[-1ex]
& \hspace{17ex} + \sum_{i \neq k} h_{\theta:m,i}(\theta_{i} - \hat{\theta}_{m,i}^{[j]}) + \xi_m, \nonumber
\end{align}
\vspace{-1ex}
\begin{align}
\label{eq:t_IC}
\tilde{z}_{t:m,\ell}^{[j]} &= z_{m} - \sum_{i = 1}^{K} h_{\theta:m,i}\hat{\theta}_{m,i}^{[j]} - \sum_{i \neq \ell}  h_{t:m,i}\hat{t}_{m,i}^{[j]} \\[-1ex]
&=  h_{t:m,\ell}t_{\ell} + \sum_{i = 1}^{K}  h_{\theta:m,i}(\theta_{i} - \hat{\theta}_{m,i}^{[j]}) \nonumber \\[-1ex] 
& \hspace{17ex} + \sum_{i \neq \ell}  h_{t:m,i}(t_{i} - \hat{t}_{m,i}^{[j]}) + \xi_m. \nonumber
\end{align}
\end{subequations}

Thus, the conditional \ac{PDF} of the \ac{sIC} symbols are defined as
\begin{subequations}
\begin{eqnarray}
\label{eq:del_t_cond_PDF}
&\!\!\!\!p_{\tilde{\mathrm{z}}_{\theta:m,k}^{[j]} \mid \mathrm{\theta}_{k}}(\tilde{z}_{\theta:m,k}^{[j]}|\theta_{k}) \propto \mathrm{exp}\bigg[ -\frac{|\tilde{z}_{\theta:m,k}^{[j]} - h_{\theta:m,k} \theta_{k}|^2}{\sigma_{\theta:m,k}^{2\,[j]}} \bigg],&\\
&\!\!\!\!p_{\tilde{\mathrm{z}}_{t:m,\ell}^{[j]} \mid \mathrm{t}_{\ell}}(\tilde{z}_{t:m,\ell}^{[j]}|t_{\ell}) \propto \mathrm{exp}\bigg[ -\frac{|\tilde{z}_{t:m,\ell}^{[j]} - h_{t:m,\ell} t_{\ell}|^2}{\sigma_{t:m,\ell}^{2\,[j]}} \bigg],&
\end{eqnarray}
\end{subequations}
with the corresponding conditional variances{\color{black}, incorporating the composite noise power $N_0$,} given by
\begin{subequations}
\label{eq:del_t_theta_var}
\begin{equation}
\label{eq:del_theta_var}
\sigma_{\theta:m,k}^{2\,[j]}\! = \!\displaystyle\sum\limits_{i \neq k} \big|h_{\theta:m,i}\big|^2\psi_{\theta:m,i}^{[j]}\! +\! \sum\limits_{i = 1}^{K} \big|h_{t:m,i}\big|^2\psi_{t:m,i}^{[j]} \!+\! N_{0} \in \mathbb{R},
\end{equation}
\begin{equation}
\label{eq:del_t_var}
\sigma_{t:m,\ell}^{2\,[j]} \!=\! \displaystyle\sum\limits_{i = 1}^{K} \big|h_{\theta:m,i}\big|^2\psi_{\theta:m,i}^{[j]}\! +\! \sum\limits_{i \neq \ell} \big|h_{t:m,i}\big|^2\psi_{t:m,i}^{[j]}\! +\! N_{0} \in \mathbb{R},
\end{equation}
where the corresponding \ac{MSE} are defined as 
%
\begin{equation}
  \psi_{\theta:m,k}^{[j]} = \mathbb{E}_{\mathsf{\theta}_k}\!\big[ | \theta_{k} - \hat{\theta}_{m,k}^{[j]} |^2 \big],
\end{equation}
\begin{equation}
  \psi_{t:m,\ell}^{[j]} = \mathbb{E}_{\mathsf{t}_{\ell}}\!\big[ | t_{\ell} - \hat{t}_{m,\ell}^{[j]} |^2 \big].
\end{equation}
\end{subequations}

Since the conditional \ac{PDF} is known, the extrinsic \ac{PDF} can be found next via \vspace{-1.5ex}
\begin{equation}
\begin{aligned}
\prod_{i \neq m} p_{\tilde{\mathsf{z}}_{\theta:i,k}^{[j]} \mid \mathsf{\theta}_{k}}\left(\tilde{z}_{\theta:i,k}^{[j]} \mid \theta_{k}\right) &\propto \mathrm{exp}\bigg[ -\frac{|\theta_{k} - \bar{\theta}_{m,k}^{[j]}|^2}{\bar{v}_{\theta:m,k}^{[j]}} \bigg], \\
\prod_{i \neq m} p_{\tilde{\mathsf{z}}_{t:i,\ell}^{[j]} \mid \mathrm{t}_{\ell}}\left(\tilde{z}_{t:i,\ell}^{[j]} \mid t_{\ell}\right) &\propto \mathrm{exp}\bigg[ -\frac{|t_{\ell} - \bar{t}_{m,\ell}^{[j]}|^2}{\bar{v}_{t:m,\ell}^{[j]}} \bigg],
\end{aligned}
\label{eq:del_t_extr_PDF}
\end{equation}
with the extrinsic means and variances are defined as
\begin{subequations}
\label{eq:del_t_theta_extr_mean}
\begin{align}
\bar{\theta}_{m,k}^{[j]} &= \bar{v}_{\theta:m,k}^{[j]} \bigg( \sum_{i \neq m} \frac{h_{\theta:i,k} \cdot \tilde{z}_{\theta:i,k}^{[j]}}{ \big(\sigma_{\theta:i,k}^{[j]}\big)^2} \bigg)\in \mathbb{R}, \label{eq:del_the_extr_mean}\\[1ex]
\bar{t}_{m,\ell}^{[j]} &= \bar{v}_{t:m,\ell}^{[j]} \bigg( \sum_{i \neq m} \frac{h_{t:i,\ell} \cdot \tilde{z}_{t:i,\ell}^{[j]}}{ \big(\sigma_{t:i,\ell}^{[j]}\big)^2} \bigg)\in \mathbb{R}, \label{eq:del_t_extr_mean}
\end{align}
\end{subequations}

\begin{subequations}
\label{eq:del_t_theta_extr_var}
\begin{align}
\bar{v}_{\theta:m,k}^{[j]} &= \bigg( \sum_{i \neq m} \frac{|h_{\theta:i,k}|^2}{\big(\sigma_{\theta:i,k}^{[j]}\big)^2} \bigg)^{\!\!\!-1}  \!\!\!\!\in \mathbb{R}, \label{eq:del_the_extr_var} \\[1ex]
\bar{v}_{t:m,\ell}^{[j]} &= \bigg( \sum_{i \neq m} \frac{|h_{t:i,\ell}|^2}{\sigma_{t:m,\ell}^{2\,[j]}} \bigg)^{\!\!\!-1} \!\!\!\!\in \mathbb{R}. \label{eq:del_t_extr_var}
\end{align}
\end{subequations}

Finally{\color{black},} the denoisers with a Gaussian prior are given by 
\begin{subequations}
\begin{equation}
\label{eq:del_t_soft_est}
\begin{aligned}
\check{\theta}_{m,k} &= \frac{\phi_{\theta} \cdot \bar{\theta}_{m,k}^{[j]}}{\phi_{\theta} + \bar{v}_{\theta:m,k}^{[j]}} \in \mathbb{R}, 
& \check{t}_{m,\ell} &= \frac{\phi_{t} \cdot \bar{t}_{m,\ell}^{[j]}}{\phi_{t} + \bar{v}_{t:m,\ell}^{[j]}} \in \mathbb{R}, \\
\end{aligned}
\end{equation}
\begin{equation}
\label{eq:del_t_est_mse}
\hspace{-2ex}\check{\psi}_{\theta:m,k} = \frac{\phi_{\theta} \cdot \bar{v}_{\theta:m,k}^{[j]}}{\phi_{\theta}\! +\! \bar{v}_{\theta:m,k}^{[j]}} \in \mathbb{R},\;
 \check{\psi}_{t:m,\ell} = \frac{\phi_{t} \cdot \bar{v}_{t:m,\ell}^{[j]}}{\phi_{t}\! +\! \bar{v}_{t:m,\ell}^{[j]}} \in \mathbb{R}, 
\end{equation}
where $\phi_{\theta}$ and $\phi_{t}$ are the variance of the individual elements in $\boldsymbol{\theta}$ and $\boldsymbol{t}$.
\end{subequations}

Subsequently, the soft-replicas are iteratively updated via \vspace{-2ex}
\begin{subequations}
\begin{align}
\hat{x}_{m, k}^{[j+1]} = \rho \hat{x}_{m, k}^{[j]} + (1 - \rho) \check{x}_{m, k}^{[j]}, \\[1ex]
\psi_{m, k}^{[j+1]} = \rho {\psi}_{m, k}^{[j]} + (1 - \rho) \check{\psi}_{m, k}^{[j]},
\end{align}
\label{eq:damped_update}
\end{subequations}
where $\rho$ is the damping factor, while the algorithm performs $j_\mathrm{max}$ iterations of the message passing method or until a convergence criteria is met, after which the consensus estimates are obtained as 
\begin{subequations}
\label{eq:del_t_final_est}
\begin{eqnarray}
&\tilde{\theta}_{k} = \!\!\bigg( \sum\limits_{m = 1}^{M} \frac{|h_{\theta:m,k}|^2}{\big(\sigma_{\theta:m,k}^{[j_\mathrm{max}]}\big)^2} \bigg)^{\!\!\!-1} \! \! \bigg( \sum\limits_{m = 1}^{M} \frac{h_{\theta:m,k} \cdot \tilde{z}_{\theta:m,k}^{[j_\mathrm{max}]}}{ \big(\sigma_{\theta:m,k}^{[j_\mathrm{max}]}\big)^2} \bigg) \in \mathbb{R},\;\;\;& \label{eq:del_t_final_est_theta} \\
&\tilde{t}_{\ell} =\!\! \bigg( \sum\limits_{m = 1}^{M} \frac{|h_{t:m,\ell}|^2}{\big(\sigma_{t:m,\ell}^{[j_\mathrm{max}]}\big)^2} \bigg)^{\!\!\!-1} \!\bigg( \sum\limits_{m = 1}^{M} \frac{h_{t:m,\ell} \cdot \tilde{z}_{t:m,\ell}^{[j_\mathrm{max}]}}{\big(\sigma_{t:m,\ell}^{[j_\mathrm{max}]}\big)^2} \,\bigg) \in \mathbb{R}.\;\;\;&
\label{eq:del_t_final_est_t}
\end{eqnarray}
\label{eq:del_final_est}
\end{subequations}

Even though the message passing rules described by equations \eqref{eq:del_t_soft_IC}-\eqref{eq:del_t_final_est} are complete to yield the estimated rotation angles and translation vectors, there is a difference in effective channel powers of $\boldsymbol{H}_{\theta}$ and $\boldsymbol{H}_{t}$ in \eqref{eq:delta_t_lin_syst}, where the latter is typically much larger, resulting from the absolute positions of the anchors and landmark points, which lead to good estimates of the translation vector elements, but poor estimates of the rotation angles in a joint estimation described by the \ac{GaBP} procedure.

\begin{algorithm}[t]
\caption[]{: Double \ac{GaBP} for \ac{RBL} Parameter Estimation}
\label{alg:RBL_GaBP}
\hspace*{\algorithmicindent}
\begin{algorithmic}[1]
\vspace{-0.9ex}
\Statex \hspace{-4ex} \textbf{Input:} $\boldsymbol{z}_n \!~\forall n,  \boldsymbol{H}_{\theta}, \boldsymbol{H}_{t}, \phi_{\theta}, \phi_{t}, N_0, j_\mathrm{max}, \rho$. \vspace{-1.25ex}
\Statex \hspace{-4.4ex} \hrulefill
\Statex \hspace{-4ex}  \textbf{Output:} $\tilde{\theta}_{k}$ and $\tilde{t}_{\ell} ~\forall k,\ell$ (for all sensor nodes $\forall n$); \vspace{-1.5ex}
\Statex \hspace{-4.4ex} \hrulefill 
\Statex \hspace{-3.2ex} \textit{{Perform}} $\forall n, m, k, \ell:$ \vspace{0.25ex}
\State {\color{black}Initialize} $\hat{\theta}_{m, k}^{[1]}$,  $\hat{t}_{m, \ell}^{[1]}$, $\psi_{\theta:m, k}^{[1]}$, $\psi_{t:m, \ell}^{[1]}$;
\For {$j = 1$ to $j_\mathrm{max}$}
\State \hspace{-3.5ex} Compute \ac{sIC} symbols $\tilde{z}_{\theta:m,k}^{[j]}$, $\tilde{z}_{t:m,\ell}^{[j]}$ via \eqref{eq:del_t_soft_IC};
\State \hspace{-3.5ex} Compute conditional variances $\sigma_{\theta:m,k}^{2\,[j]}, \sigma_{t:m,\ell}^{2\,[j]}$ via \eqref{eq:del_t_theta_var};
\State \hspace{-3.5ex} Compute extrinsic means $\bar{\theta}_{m,k}^{[j]}$, $\bar{t}_{m,\ell}^{[j]}$ via \eqref{eq:del_t_theta_extr_mean};
\State \hspace{-3.5ex} Compute extrinsic variances $\bar{v}_{\theta:m,k}^{[j]}$, $\bar{v}_{t:m,\ell}^{[j]}$ via \eqref{eq:del_t_theta_extr_var}; \vspace{-0.5ex}
\State \hspace{-3.5ex} Denoise the beliefs $\check{\theta}_{m,k}, \check{t}_{m,\ell}$ via \eqref{eq:del_t_soft_est}; \vphantom{ $\check{x}_{m, k}^{[j]}$}  \vspace{-0.5ex}
\State \hspace{-3.5ex} Denoise the error variances $\check{\psi}_{\theta:m,k} \check{\psi}_{t:m,\ell}$ via \eqref{eq:del_t_est_mse}; \vphantom{ $\check{x}_{m, k}^{[j]}$}  \vspace{-0.5ex}
\State \hspace{-3.5ex} Update the soft-replicas with damping; \vphantom{ $\check{x}_{m, k}^{[j]}$} 
\State \hspace{-3.5ex} Update the eff. rotation channel matrix $\boldsymbol{H}_{\theta}$ via \eqref{H_new};%
\EndFor
\State Obtain final consensus estimates $\tilde{\theta}_{k}, \tilde{t}_{\ell}$ via \eqref{eq:del_t_final_est};
\State Obtain interference-cancelled system via \eqref{eq:new_lin_syst};
\For {$j = 1$ to $j_\mathrm{max}$}  \vspace{-0.2ex}
\State \hspace{-3.5ex} Compute \ac{sIC} symbols $\tilde{z}{'}_{\!\!\theta:m,k}^{[j]}$ via \eqref{eq:new_lin_SIC};  \vspace{-0.5ex}
\State \hspace{-3.5ex} Compute conditional variances $\sigma_{\theta:m,k}^{2\,[j]}$ via \eqref{eq:new_lin_condvar};  \vspace{-0.5ex}
\State \hspace{-3.5ex} Compute extrinsic means $\bar{\theta}_{m,k}^{[j]}$ via \eqref{eq:del_the_extr_mean};  \vspace{-0.5ex}
\State \hspace{-3.5ex} Compute extrinsic variances $\bar{v}_{\theta:m,k}^{[j]}$ via \eqref{eq:del_the_extr_var};  \vspace{-0.75ex}
\State \hspace{-3.5ex} Denoise the beliefs $\check{\theta}_{m,k}$ via \eqref{eq:del_t_soft_est};  \vspace{-0.5ex} \vphantom{ $\check{x}_{m, k}^{[j]}$}
\State \hspace{-3.5ex} Denoise the error variances $\check{\psi}_{\theta:m,k}$ via \eqref{eq:del_t_est_mse}; \vphantom{ $\check{x}_{m, k}^{[j]}$} \vspace{-0.75ex}
\State \hspace{-3.5ex} Update the soft-replicas with damping; \vphantom{ $\check{x}_{m, k}^{[j]}$} 
\State \hspace{-3.5ex} Update the eff. rotation channel matrix $\boldsymbol{H}_{\theta}$ via \eqref{H_new};
\EndFor
\State Obtain refined consensus estimates {\color{black}$\tilde{\theta}_{k}, \tilde{t}_{\ell}$ via \eqref{eq:del_final_est}};
\end{algorithmic} 
\hspace*{\algorithmicindent}
\vspace{-2ex}
\end{algorithm}

Such an effect can be intuitively explained by the fact that a small change in the rotation angles leads to a small change in the sensor positions, while a small change in the translation vector leads to a large change in the sensor positions, as visualized in Fig.~\ref{fig:RB_trans_plot}.

In light of the above, to counter the problem of the aforementioned error behavior of the rotation angle parameters $\boldsymbol{\theta}$, we propose an interference cancellation-based approach to remove the components corresponding to the translation of the rigid body, and perform the \ac{GaBP} again only on the rotation angle parameters.
By using the estimated consensus translation vector $\tilde{\boldsymbol{t}} \triangleq [\tilde{t}_1, \tilde{t}_2, \tilde{t}_3]^\intercal \!\in \mathbb{R}^{3\times 1}$ obtained at the end of the \ac{GaBP} via \eqref{eq:del_t_final_est_t}, the interference-cancelled system is given by
\begin{equation}
\label{eq:new_lin_syst}
\boldsymbol{z}_{n}' \triangleq \boldsymbol{z}_{n} - \boldsymbol{H}_{t} \tilde{\boldsymbol{t}} = \boldsymbol{H}_{\theta} \boldsymbol{\theta} + \boldsymbol{\xi}_{n} \in \mathbb{R}^{M \times 1}.
\end{equation}

Finally, the \ac{GaBP} procedure to estimate the rotation angle parameters $\boldsymbol{\theta}$ is similar to the previous steps, except for the factor node equations, which yields
\begin{subequations}
\begin{equation}
\label{eq:new_lin_SIC}
\tilde{z}{'}_{\!\!\theta:m,k}^{[j]} = z'_{m} - \sum_{i \neq k} h_{\theta:m,i}\hat{\theta}_{m,i}^{[j]} \in \mathbb{R},
\end{equation}
\begin{equation}
\label{eq:new_lin_condvar}
\sigma_{\theta:m,k}^{2\,[j]} = \sum_{i \neq k} \big|h_{\theta:m,i}\big|^2\psi_{\theta:m,i}^{[j]} + N_{0} \in \mathbb{R}.
\end{equation}
\end{subequations}

The second loop to refine the rotation estimates is concatenated with the bivariate \ac{GaBP} to describe the complete estimation process of the rigid body parameters $\boldsymbol{\theta}$ and $\boldsymbol{t}$ {\color{black}per landmark point,} as summarized by Algorithm \ref{alg:RBL_GaBP}.

{\color{black}
The final rigid body parameter estimates are then obtained via average combination of the $N$ per-landmark consensus estimates from Algorithm \ref{alg:RBL_GaBP}, $i.e.$, $\tilde{\theta}_{k}^{(n)}$ and $\tilde{t}_{l}^{(n)}$, given by
\begin{subequations}
\label{eq:landmark_avg}
\begin{eqnarray}
\tilde{\boldsymbol{\theta}} = \frac{1}{N}\sum_{n=1}^{N} \Big[\tilde{\theta}_{1}^{(n)}, \tilde{\theta}_{2}^{(n)}, \ldots, \tilde{\theta}_{K}^{(n)}\Big]^\intercal \in \mathbb{R}^{K\times1}, \label{eq:landmark_avg_theta}\\
\tilde{\boldsymbol{t}} = \frac{1}{N}\sum_{n=1}^{N} \Big[\tilde{t}_{1}^{(n)}, \tilde{t}_{2}^{(n)}, \ldots, \tilde{t}_{K}^{(n)}\Big]^\intercal \in \mathbb{R}^{K\times1}. \label{eq:landmark_avg_t}
\end{eqnarray}
\end{subequations}
}

\vspace{-2ex}
\section{Performance Evaluation}
\label{sec:res}

To provide a comprehensive performance evaluation of the proposed \ac{GaBP}-based \ac{RBL} algorithm, we compare it against the \ac{SotA} method described in \cite{fuehrling20246drigidbodylocalization}, which also uses \ac{GaBP} for the estimation of the rigid body transformation parameters, but does not use the proposed quadratic approximation of the rotation matrix, and instead relies on the small-angle approximation{\color{black}, as well as the \ac{LS} method of \cite{Chen_2015} and a \ac{SDR}-based approach proposed in \cite{Jiang2018}}. 
{\color{black}Since the proposed and \ac{SotA} \ac{GaBP} schemes are otherwise algorithmically identical, this comparison also constitutes a direct ablation of the proposed angle approximation itself.}
The simulation parameters are illustrated in Fig.~\ref{fig:RB_anchor}, where the rigid body is composed of $N$ landmark points, whose positions are known in the local coordinate system of the rigid body, and $M$ anchors with known absolute positions in the global coordinate system. 
The conformation matrix of the target rigid body is given by
\begin{equation*}\label{eq:C_mat}
\boldsymbol{C} \!=\!\!
\resizebox{0.39 \textwidth}{!}{$\left[\begin{array}{lllllllc}
-0.5 &\! \phantom{-}0.5 &\! \phantom{-}0.5 &\! -0.5 &\! -0.5 &\! \phantom{-}0.5 &\! -0.5 &\! \phantom{-}0.5 \\
-0.5 &\!-0.5 &\! \phantom{-}0.5 &\!\phantom{-}0.5 &\! -0.5 &\! -0.5 & \!\phantom{-}0.5 & \!\phantom{-}0.5 \\
-0.5 &\! -0.5 &\! -0.5 &\! -0.5 & \!\phantom{-}0.5 &\! \phantom{-}0.5 &\! \phantom{-}0.5 &\! \phantom{-}0.5
\end{array}\right]
$} \!\!\in\! \mathbb{R}^{3 \times 8}\!,
\end{equation*}
while the anchors conformation matrix $\boldsymbol{A} \in \mathbb{R}^{3\times 8}$ is
\begin{equation*}\label{eq:A_mat}
\boldsymbol{A} \!=\!\!
\resizebox{0.38 \textwidth}{!}{$
\left[\begin{array}{lllllllc}
-10 &\! \phantom{-}10 &\! \phantom{-}10 &\! -10 & \!-10 & \!\phantom{-}10 &\! -10 &\! \phantom{-}10 \\
-10 & \!-10 & \!\phantom{-}10 &\!\phantom{-}10 &\! -10 & \!-10 & \!\phantom{-}10 & \!\phantom{-}10 \\
-10 & \!-10 &\! -10 &\! -10 & \!\phantom{-}10 &\! \phantom{-}10 &\! \phantom{-}10 & \!\phantom{-}10
\end{array}\right]
$}  \!\!\in\! \mathbb{R}^{3 \times 8}\!.
\end{equation*}

The rigid body parameters are generated randomly each realization, where the rotation angles $\theta_x, \theta_y, \theta_z$ follow a zero-mean Gaussian distribution of variance $\phi_{\theta} = 10 \text{deg}^2$, which is changing depending on the large or small angle scenario and the \ac{RBL} translation vector elements also follow a zero-mean Gaussian distribution of variance $\phi_{t} = 5 \text{m}^2$. 
{\color{black}Note that the proposed method and the small-angle \ac{GaBP} baseline are both initialized from, and provided with, the same prior information, such that both schemes start from an identical starting point and the reported performance difference is attributable to the angle approximation alone.}

\vspace{-2ex}

\begin{figure}[H]
\centering
{{\includegraphics[width=\columnwidth]{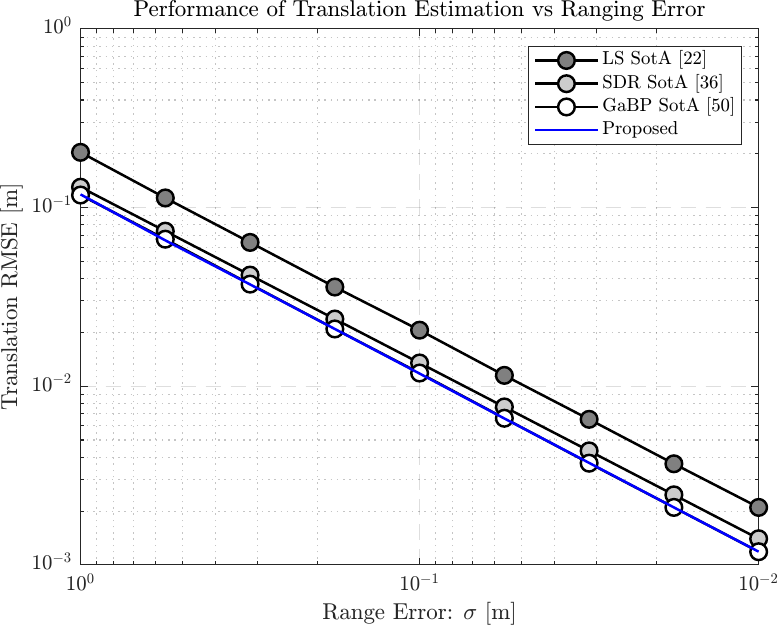}}}
\vspace{-2.5ex}
\caption[]{\color{black}\ac{RMSE} of the translation estimate of the proposed method and the \ac{SotA}, over the range error $\sigma_w$.}
\label{fig:Tra}
\end{figure}
\vspace{-7ex}
\begin{figure}[H]
\centering{{\includegraphics[width=\columnwidth]{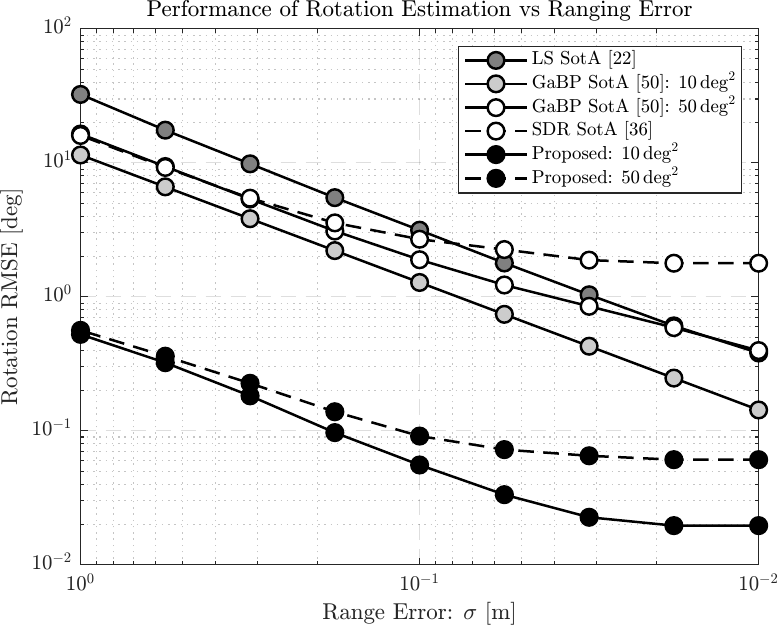}}}
\vspace{-2.5ex}
\caption[]{\color{black}\ac{RMSE} of the rotation estimate of the proposed method and the \ac{SotA}, over the range error $\sigma_w$.}
\label{fig:Rot}
\end{figure}

\vspace{-6ex}

\subsection{Numerical Results}

The first set of results in Fig.~\ref{fig:Tra} illustrates the \ac{RMSE} performance of the translation vector estimate of the proposed method, compared to the \ac{SotA} method of \cite{fuehrling20246drigidbodylocalization} that uses \ac{GaBP} to estimate the rigid body parameters, which is, however, limited by the small-angle approximation used in the system model, {\color{black}as well as the \ac{LS} and \ac{SDR} methods \cite{Chen_2015,Jiang2018}}.
It can be observed that, as expected, the translation vector estimate is equally well estimated by both {\color{black}\ac{GaBP}-based} methods, {while outperforming the \ac{LS} and \ac{SDR} methods,} over the whole error regime.
This result can be justified by the fact that the translation part of the system model is not affected by the different angle approximation used in the linearized model and thus, yields the same performance as the \ac{SotA}.

The performance is evaluated by the \ac{RMSE}, which is defined as
\begin{equation}
\mathrm{RMSE} = \sqrt{\frac{1}{E}\sum_{i=1}^{E}\|\hat{\boldsymbol{x}}^{[i]}-{\boldsymbol{x}}\|_2^2},
\end{equation}
where $\hat{\boldsymbol{x}}^{[i]}$ is the rigid body parameter vector (angle or translation) estimated during the $i$-th Monte-Carlo simulation, $\boldsymbol{x}$ is the true \ac{RBL} parameter vector, and $E = 10^4$ is the total number of independent Monte-Carlo experiments used for the analysis, and is evaluated for different noise standard deviations $\sigma_w$.
Additionally, for the \ac{GaBP} algorithms, the damping factor is set to $\rho = 0.5$, and the maximum number of iterations is set to $j_\mathrm{max} = 30$, which has empirically shown sufficient convergence. {\color{black}These values correspond to a standard hyperparameter selection for iterative message passing algorithms, and were found to provide stable convergence across all considered scenarios without per scenario tuning.}

Next, in Fig.~\ref{fig:Rot}, the \ac{RMSE} performance of the rotation angle estimate is illustrated, where again, the proposed method is compared to the {\color{black}three \ac{SotA} methods}, for different levels of the angle deviation $\phi_\theta$.
It can be observed that the general performance of the proposed method is far superior to the \ac{SotA}, especially for larger angle deviations.
While the performance of the \ac{SotA} degrades rapidly for larger angle deviations, the proposed method is robust to such changes, with an overall improved performance of more than one decade, over the whole noise regime{\color{black}, also outperforming the \ac{LS} and \ac{SDR} methods}.

\begin{figure}[H]
\centering
{{\includegraphics[width=\columnwidth]{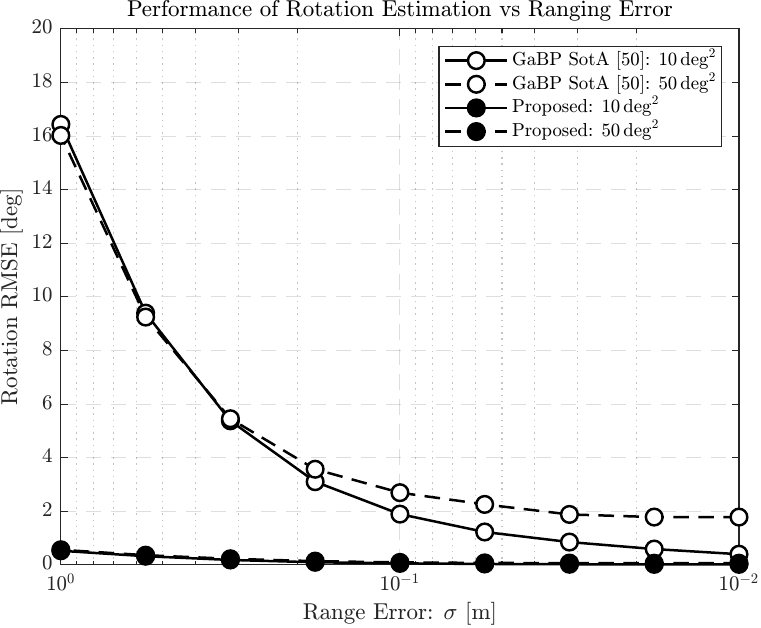}}}
\vspace{-4.5ex}
\caption[]{\color{black}\ac{RMSE} of the rotation estimate of the proposed method and the \ac{SotA}, over the range error $\sigma_w$.}
\label{fig:Rot_lin}
\end{figure}

Finally, since the gain in terms of robustness is not clearly visible in the logarithmic scale of Fig.~\ref{fig:Rot}, in Fig.~\ref{fig:Rot_lin} the results {\color{black} of the \ac{GaBP}-based approaches} are presented again in a linear scale.
As can be observed, the proposed method maintains a low \ac{RMSE} even for large angle deviations, over the whole noise regime, while the \ac{SotA} method fails to provide acceptable estimates for larger angle deviations and high measurement noise.

\vspace{-4ex}

\begin{figure}[H]
\centering
{{\includegraphics[width=\columnwidth]{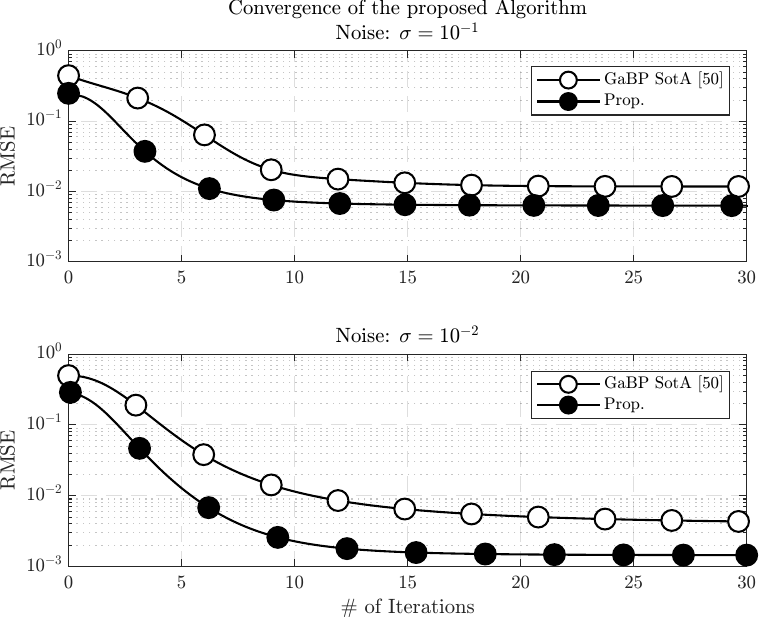}}}
\vspace{-2.5ex}
\caption{\color{black}\ac{RMSE} convergence of the estimates of the proposed method and the \ac{SotA}, over the algorithmic iterations, for $\sigma_w = 10^{-1}$ and $\sigma_w = 10^{-2}$.}
\label{fig:convergence}
\end{figure}
\vspace{-2ex}

\subsection{Complexity {\color{black} and Convergence} Analysis}

Next, the convergence behavior of the proposed method is analyzed, and compared to the \ac{SotA} method of \cite{fuehrling20246drigidbodylocalization}, which also uses \ac{GaBP} for the estimation of the rigid body transformation parameters{\color{black}, since the \ac{LS} and \ac{SDR} approaches are not iterative}.
It can be observed in Fig.~\ref{fig:convergence} that the proposed method converges to a lower \ac{RMSE} than the \ac{SotA}, as expected from the previous results, while the proposed method converges slightly faster than the \ac{SotA} method.
As can be observed, the proposed method converges in less than $10$ iterations, while the \ac{SotA} method converges in around $15$ iterations, which is similar for all noise levels $\sigma_w$, but only shown for $\sigma_w = 10^{-1}$ {\color{black} and $\sigma_w = 10^{-2}$} for brevity.
{\color{black}
Note that the smooth and consistent \ac{RMSE} trends observed across the full range of noise levels and angle deviations considered in Figs.~\ref{fig:Tra}-\ref{fig:Rot_lin}, all computed using the same fixed \ac{GaBP} hyperparameters ($j_\mathrm{max}=30$, $\rho=0.5$) as in Fig.~\ref{fig:convergence}, provide broader empirical evidence of reliable convergence beyond the single operating point highlighted here. 
We further note that a formal convergence proof is not established in this work, since the effective channel matrix $\boldsymbol{H}_\theta$ is relinearized at every iteration based on the previous angle estimate.
The resulting procedure differs from the static \ac{GaBP} setting for which sufficient convergence conditions such as walk-summability or diagonal dominance of the precision matrix are known \cite{Malioutov2006,WeissFreeman2001,SuWu2015}.
A theoretical convergence analysis of this iteratively-relinearized setting is left for future work to establish.
}

In addition to the performance {\color{black} and convergence} analysis, for the sake of completeness, the computational complexity of the proposed methods are compared to the \ac{SotA} method of \cite{fuehrling20246drigidbodylocalization}, given in big-$O$ notation.
While the proposed method clearly outperforms the \ac{SotA} methods in terms of performance, its complexity in big-$O$ notation is identical, with $O(NMK^2)$ for both \ac{GaBP} algorithms, where $N$ is the number of landmark points, $M$ is the number of anchors, and $K$ is the dimension of space.
The complexity compared to the \ac{SotA} can be well justified, since the only difference is the iterative update of the effective channel matrix $\boldsymbol{H}_\theta$ in \eqref{H_new}, which does not change the overall complexity order.
Finally, in terms of convergence behavior, the convergence of the rotation matrix estimate of the proposed method converges in a similar manner as the \ac{SotA}, but to a lower \ac{RMSE}, with the proposed method converging in less than $10$ iterations and the \ac{SotA} in around $15$ iterations across the considered noise levels $\sigma_w$, consistent with Fig.~\ref{fig:convergence}, and thus, a dedicated convergence figure for the rotation estimate is omitted for brevity.

\begin{figure*}[t!]
\setcounter{equation}{29}
\normalsize
\begin{equation}
  \label{eq:Approx_Q}
    \bm{Q}^{[i]}\!\!=\!\!\text{\scalebox{1}{$
\left[
\begin{array}{@{\,}c@{\;\,}c@{\;\,}c@{\,}}
\gamma^{2}
-\gamma\delta
\bigl(\theta_{y}^{[i-1]}\theta_{y}^{[i]}+\theta_{z}^{[i-1]}\theta_{z}^{[i]}\bigr)
&
-\gamma\beta\,\theta_{z}^{[i]}
\;-\;
\gamma\alpha\,\theta_{z}^{[i-1]}\theta_{z}^{[i]}
&
\beta\,\theta_{y}^{[i]}
\;+\;
\alpha\,\theta_{y}^{[i-1]}\theta_{y}^{[i]}
\\[10pt]
\gamma\beta\,\theta_{z}^{[i]}
\;+\;
\gamma\alpha\,\theta_{z}^{[i-1]}\theta_{z}^{[i]}
&
\gamma^{2}
-\gamma\delta
\bigl(\theta_{x}^{[i-1]}\theta_{x}^{[i]}+\theta_{z}^{[i-1]}\theta_{z}^{[i]}\bigr)
&
-\gamma\beta\,\theta_{x}^{[i]}
\;-\;
\gamma\alpha\,\theta_{x}^{[i-1]}\theta_{x}^{[i]}
\\[10pt]
-\beta\,\theta_{y}^{[i]}
\;-\;
\alpha\,\theta_{y}^{[i-1]}\theta_{y}^{[i]}
&
\gamma\beta\,\theta_{x}^{[i]}
\;+\;
\gamma\alpha\,\theta_{x}^{[i-1]}\theta_{x}^{[i]}
&
\gamma^{2}
-\gamma\delta
\bigl(\theta_{x}^{[i-1]}\theta_{x}^{[i]}+\theta_{y}^{[i-1]}\theta_{y}^{[i]}\bigr)\\
\end{array}\right]$}},
\end{equation}
\setcounter{equation}{29}
\hrulefill
\vspace{-3ex}
\end{figure*}

\section{Conclusion}
We proposed a novel \ac{RBL} algorithm, capable of estimating the rigid body transformation parameters, $i.e.$, translation and rotation, with high accuracy for both small and large angle deviations, based on range measurements that utilizes a novel quadratic approximation of the rotation matrix and a double \ac{GaBP}-based estimation algorithm.
First, the conventional rigid body system model is linearized by using a novel quadratic approximation of the rotation matrix, which is valid for both small and large angle deviations.

Subsequently, a double \ac{GaBP}-based estimation algorithm is proposed, employing the novel linearized system model to jointly estimate the rigid body transformation parameters.
Simulation results illustrate the good performance of the proposed technique in terms of \ac{RMSE} as a function of the measurement error, outperforming the \ac{SotA} in all error regimes for the rotation estimate, with great accuracy for large angle deviations, while preserving the same computational complexity and convergence behavior, as well as the same performance for the translation estimation.
{\color{black}
Future work will focus on the extension of the proposed method to make the method robust to even larger angle deviations, as well as to add further proof of convergence in the iteratively-relinearized \ac{GaBP} setting and a more detailed numerical evaluation with respect to the \ac{GaBP} hyperparameters.
}
\appendices

\section{Linear Model Derivation}
\label{app:Derivation}

In order to prove that \eqref{eq:quad_approx} is indeed equivalent to the vectorized version of the rotation matrix $\mathbf{Q}$ given by \eqref{eq:rotation_matrix}, the formulation can be derived, starting from \eqref{eq:rotation_matrix}, applying the corresponding approximation.

Next, all the pairs of sine and cosine functions are approximated by the proposed quadratic approximation given by \eqref{eq:quad_approx}, which lead to the following approximations, written as
\begin{subequations}
  \allowdisplaybreaks
  \begin{align}
\cos(\theta_z)\sin(\theta_y) 
&\approx \gamma\beta\, \theta_y^{[i]} + \gamma\alpha\, \theta_y^{[i-1]} \theta_y^{[i]}, \\
\cos(\theta_x)\sin(\theta_z) 
&\approx \gamma\beta\, \theta_z^{[i]} + \gamma\alpha\, \theta_z^{[i-1]} \theta_z^{[i]}, \\
\cos(\theta_z)\sin(\theta_x) 
&\approx \gamma\beta\, \theta_x^{[i]} + \gamma\alpha\, \theta_x^{[i-1]} \theta_x^{[i]}, \\
\cos(\theta_x)\sin(\theta_y) 
&\approx \gamma\beta\, \theta_y^{[i]} + \gamma\alpha\, \theta_y^{[i-1]} \theta_y^{[i]}, \\
\cos(\theta_y)\sin(\theta_x) 
&\approx \gamma\beta\, \theta_x^{[i]} + \gamma\alpha\, \theta_x^{[i-1]} \theta_x^{[i]}, \\
\cos(\theta_x)\cos(\theta_z) 
&\approx \gamma^2 
      - \gamma\delta \left( \theta_x^{[i-1]} \theta_x^{[i]} + \theta_z^{[i-1]} \theta_z^{[i]} \right), \\
\cos(\theta_x)\cos(\theta_y) 
&\approx \gamma^2 
      - \gamma\delta \left( \theta_x^{[i-1]} \theta_x^{[i]} + \theta_y^{[i-1]} \theta_y^{[i]} \right),
      \label{eq:approx_pairs}
\end{align}
\end{subequations}
{\color{black}which, while not exhaustive, are the most relevant terms for the derivation of the linearized system model that lead to a highly accurate approximation of the rotation matrix, as shown in Fig.~\ref{fig:ApproxERR}.}

Next, with the approximation terms in hand, all the terms are substituted into the original rotation matrix $\mathbf{Q}$, as defined in \eqref{eq:rotation_matrix}.
The result of said substitution is described in \eqref{eq:Approx_Q} above\footnote{\color{black}Note that while the original rotation matrix is part of the $SO(3)$ group, this property does not hold for approximated matrices. Nevertheless, since the estimated parameters are the yaw, pitch, and roll angles, a final matrix satisfying the orthogonality constraint can be obtained.} and can then be vectorized to yield the final result given by \eqref{eq:quad_approx}, described in compact matrix form, whereas the full vectorized form is given as
\begin{equation}
\begin{split}
\operatorname{vec}\bigl(\mathbf{Q}^{[i]}\bigr)&=
\boldsymbol{\gamma}
+
\bigl(\beta\,\mathbf L
      +\alpha\,\mathbf L\boldsymbol\Theta
      -\delta\,\mathbf D\boldsymbol\Theta\bigr)\,
      \boldsymbol{\theta}^{[i]}
\\&=
\begin{bmatrix}
\gamma^{2}
-\gamma\delta \bigl(\theta_{y}^{[i-1]}\theta_{y}^{[i]}+\theta_{z}^{[i-1]}\theta_{z}^{[i]}\bigr)
\\[0.2ex]
\gamma\beta\,\theta_{z}^{[i]}
+\gamma\alpha\,\theta_{z}^{[i-1]}\theta_{z}^{[i]}
\\[0.2ex]
-\beta\,\theta_{y}^{[i]}
-\alpha\,\theta_{y}^{[i-1]}\theta_{y}^{[i]}
\\[0.2ex]
-\gamma\beta\,\theta_{z}^{[i]}
-\gamma\alpha\,\theta_{z}^{[i-1]}\theta_{z}^{[i]}
\\[0.2ex]
\gamma^{2}
-\gamma\delta \bigl(\theta_{x}^{[i-1]}\theta_{x}^{[i]}+\theta_{z}^{[i-1]}\theta_{z}^{[i]}\bigr)
\\[0.2ex]
\gamma\beta\,\theta_{x}^{[i]}
+\gamma\alpha\,\theta_{x}^{[i-1]}\theta_{x}^{[i]}
\\[0.2ex]
\beta\,\theta_{y}^{[i]}
+\alpha\,\theta_{y}^{[i-1]}\theta_{y}^{[i]}
\\[0.2ex]
-\gamma\beta\,\theta_{x}^{[i]}
-\gamma\alpha\,\theta_{x}^{[i-1]}\theta_{x}^{[i]}
\\[0.2ex]
\gamma^{2}
-\gamma\delta \bigl(\theta_{x}^{[i-1]}\theta_{x}^{[i]}+\theta_{y}^{[i-1]}\theta_{y}^{[i]}\bigr)
\end{bmatrix}.
\end{split}
\end{equation}



\begin{IEEEbiography}[{\includegraphics[width=1in,height=1.25in,clip,keepaspectratio]{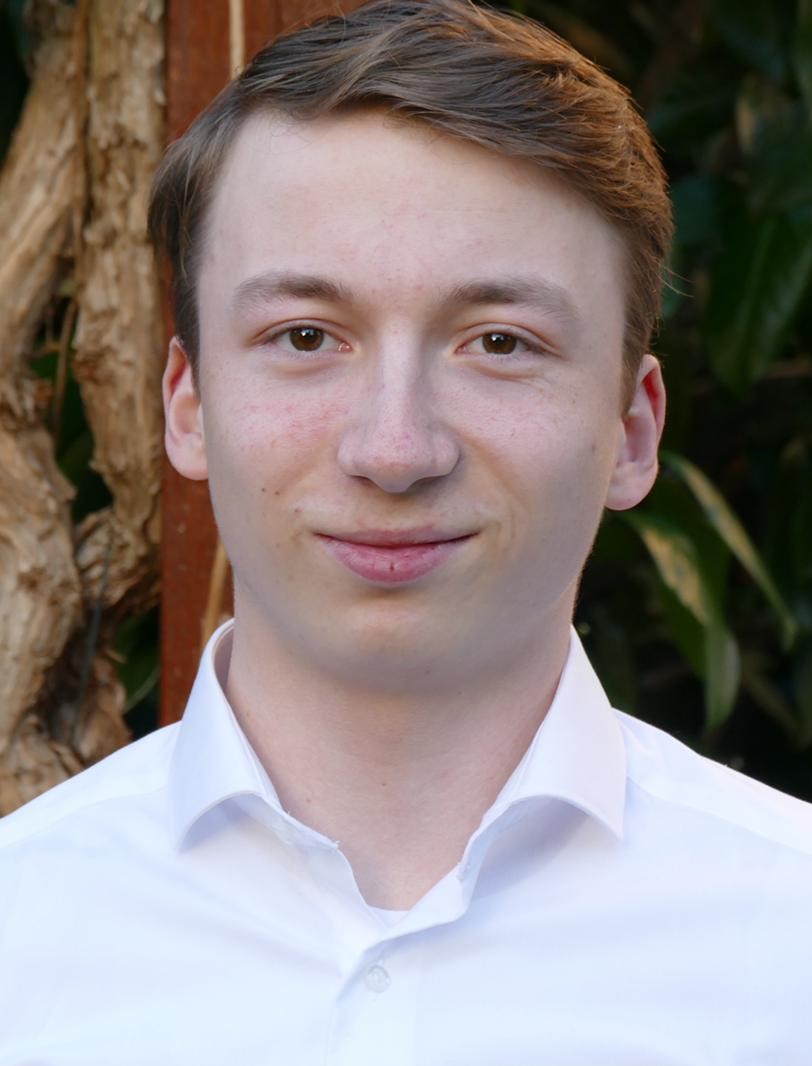}}]{Niclas F{\"u}hrling }
(Graduate Student Member, IEEE) received the B.Sc. degree in electrical and computer engineering from Jacobs University Bremen, Bremen, Germany in 2022 and the M.Sc. degree in electrical engineering with the University of Bremen in 2024, with a focus on communication and information technology. He is currently pursuing the Ph.D. degree in Electrical Engineering with Constructor University Bremen, Germany, while working on a research project  focusing on 6G connectivity. His current research interests are wireless communications, signal processing and wireless localization.
\end{IEEEbiography}
\begin{IEEEbiography}[{\includegraphics[width=1in,height=1.25in,clip,keepaspectratio]{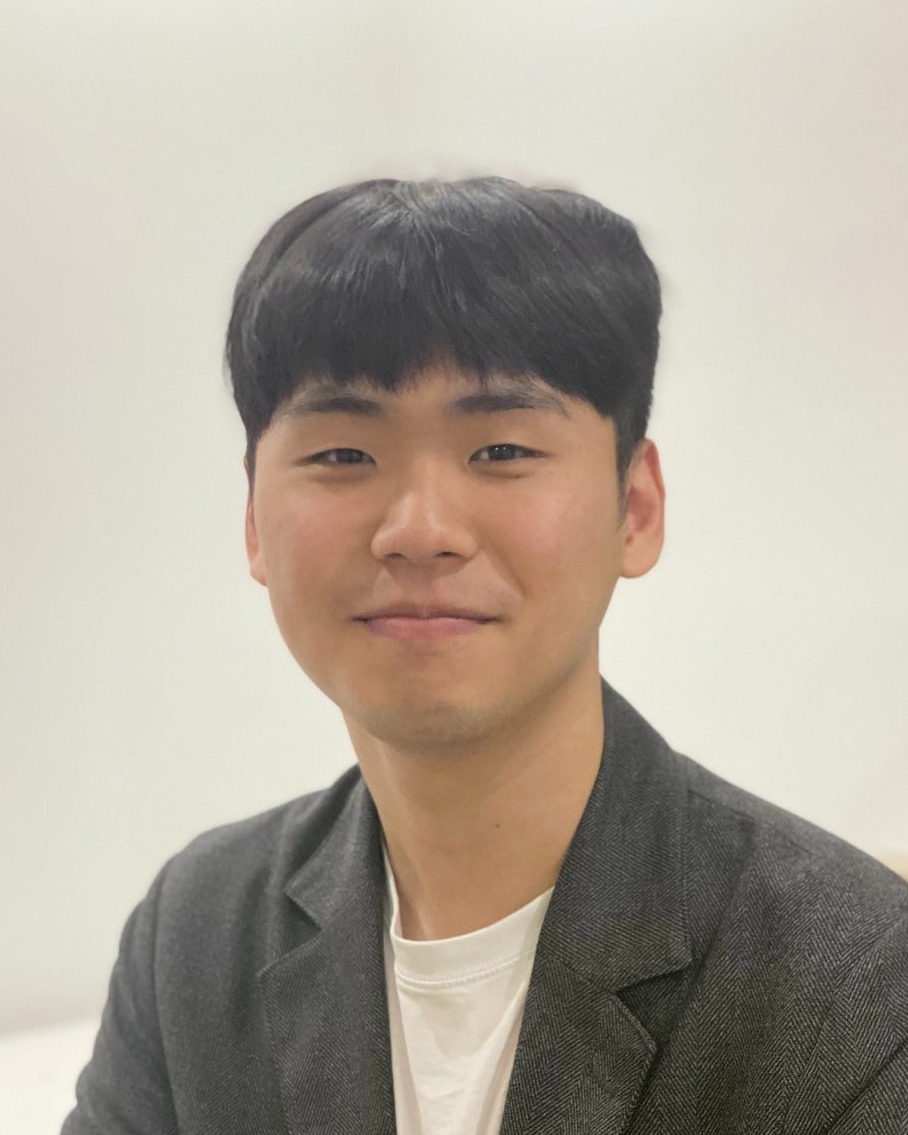}}]{Hyeon Seok Rou } (Member, IEEE) received his Ph.D. in electrical engineering from Constructor University Bremen, Germany, in 2024, and his B.Sc. in electrical and computer engineering from the same institution (formerly Jacobs University Bremen) in 2021. Since 2025, he has been a Postdoctoral Research Fellow and Lecturer at the School of Computer Science and Engineering, Constructor University Bremen, Germany. From November 2026, he will join Korea Advanced Institute of Science and Technology (KAIST)'s School of Electrical Engineering as a Research Assistant Professor under the National Research Foundation of Korea's BrainPool Fellowship.
His research interests include next-generation (B5G, 6G) wireless systems, affine frequency division multiplexing (AFDM), integrated sensing and communications (ISAC), over-the-air computing (AirComp),  high-mobility communications, multi-dimensional sparse modulation, V2X, and quantum-accelerated optimization.
In 2026, he was selected as a Brain Pool Fellow by the National Research Foundation of Korea (NRF). He has served as the chapter representative of The Korean Scientists and Engineers Association in the FRG (VeKNI e.V.) since 2021, and was the German delegate to the Young Professionals Forum (YPF) at the 3rd World Congress of Korean Scientists and Engineers in South Korea, 2025. He was a visiting researcher at the Intelligent Communications Systems Laboratory (ICL) at KAIST in 2023, and received the Korea Institute of Science and Technology Europe Research Scholarship Award in 2022. He also contributes to the IEEE Standards Association P3383 Working Group on Performance Metrics for Integrated Sensing and Communication.
\end{IEEEbiography}
\begin{IEEEbiography}[{\includegraphics[width=1in,height=1.25in,clip,keepaspectratio]{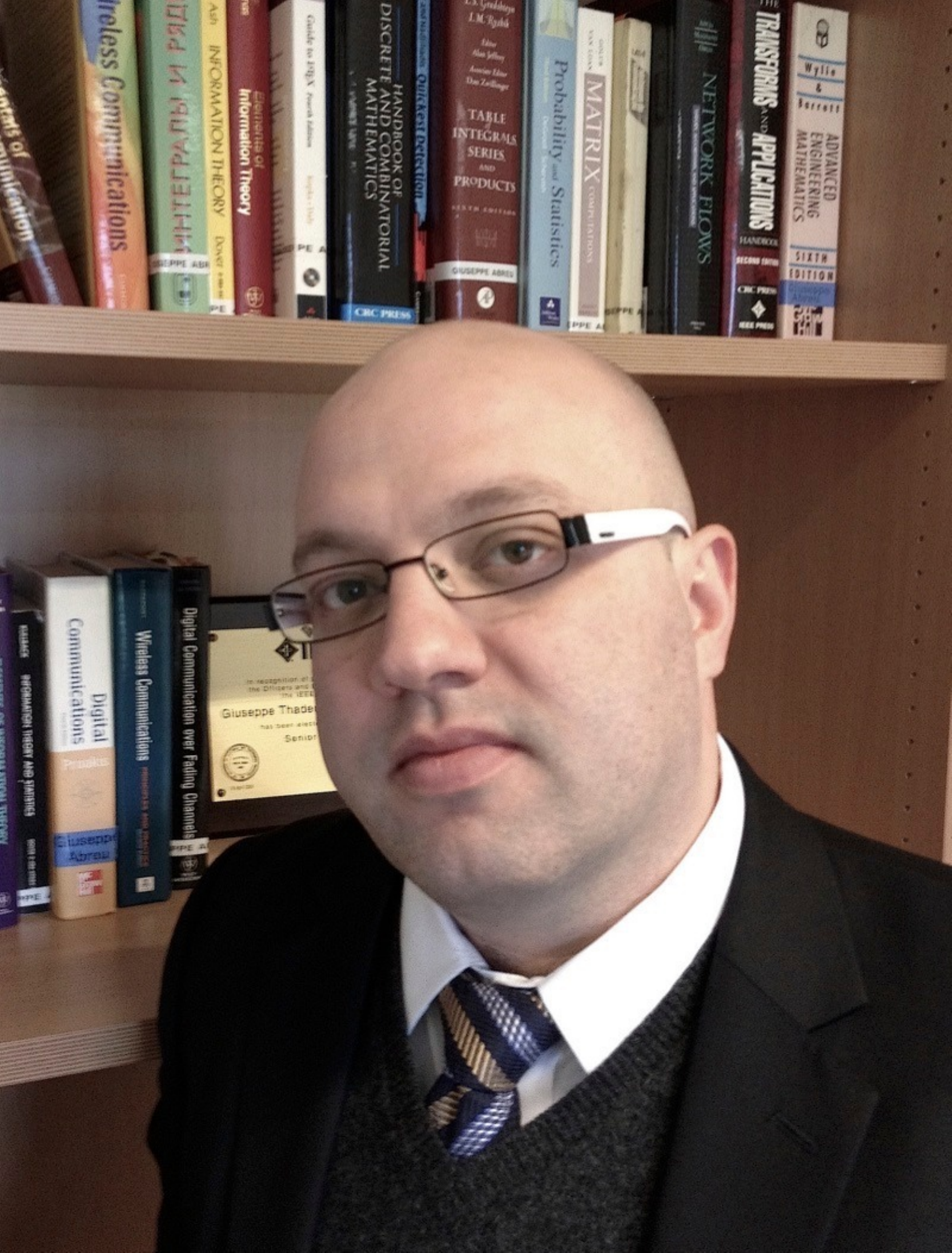}}]{Giuseppe Thadeu Freitas de Abreu } (Senior Member, IEEE) received the B.Eng. degree in electrical engineering and the specialization Latu Sensu
degree in telecommunications engineering from the
Universidade Federal da Bahia (UFBa), Salvador,
Bahia, Brazil in 1996 and 1997, respectively, and
the M.Eng. and D.Eng. degrees in physics, electrical,
and computer engineering from Yokohama National
University, Japan, in March 2001 and March 2004,
respectively.
He was a postdoctoral fellow and later
an adjunct professor (docent) in statistical signal
processing and communications theory at the Department of Electrical and
Information Engineering, University of Oulu, Finland from 2004 to 2006
and from 2006 to 2011, respectively.
Since 2011, he has been a professor
of electrical engineering at Constructor University Bremen (formerly Jacobs University, Bremen, Germany). From April 2015 to August 2018, he simultaneously held a full tenured professorship also
at the Department of Computer and Electrical Engineering, Ritsumeikan
University, Japan.
He has been a visiting scholar at Rice University (2008), Stanford University (2009), Keio University (2010), the University of Bologna (2012 and 2014) and at the KTH Royal Institute of Technology (2025). He has also been a regular visiting professor at the University of Electro-communications, Japan since 2013.
He received the Uenohara Award at Tokyo University in 2000 for his master’s thesis work and was the co-recipient of awards from various international IEEE conferences, including the Asilomar in 2009, 2012 and 2014, WPNC in 2012 and 2015, ISWCS in 2016, WPMC in 2023 and ICNC in 2025.
He was also a co-recipient of the Best Journal Paper Award by the Japanese Chapter of the IEEE Signal Processing Society in 2023 and was awarded various prestigious Fellowships from the Heiwa Nakajima Foundation, the Japanese Society for the Promotion of Science (JSPS), and twice from the National Institute for Communications Technologies (NICT), respectively in 2010, 2013, 2015 and 2018.
He has been a Visiting Scholar at Stanford University, Rice University, York University,  ETH Z{\"u}rich, over the years, and was a Scholar in Residence at the KTH Royal Institute of Technologies, Sweden, from Sep. 2025 to January 2026.
He served as an editor for various IEEE journals including the IEEE Transactions on Wireless Communications, the IEEE Transactions on Communications, the IEEE Communications Letters, the IEEE Signal Processing Letters, and the IEEE Open Journal of the Communications Society.
He was also the executive editor for the IEEE Transactions on Wireless Communications from 2017 to 2021, and since 2026 is the Editor-at-Large for Wireless Communications of the IEEE Open Journal of the Communications Society.
His research interests spans a wide range of topics on wireless communications and signal processing, including communications theory, estimation theory, optimization theory, statistical modeling, metasurfaces and electromagnetic structures, waveform design, integrated sensing and communications, over-the-air computing, wireless localization, cognitive radio, wireless security, MIMO systems, ultrawideband and millimeter wave communications, full-duplex and cognitive radio, compressive sensing, energy harvesting networks, random networks, connected vehicles networks, and many more.
\end{IEEEbiography}
\begin{IEEEbiography}[{\includegraphics[width=1in,height=1.25in,clip,keepaspectratio]{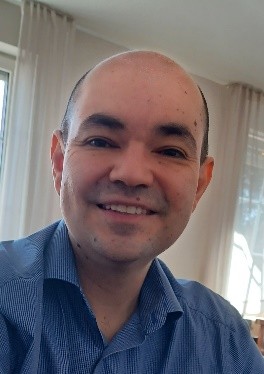}}]{David Gonz{\'a}lez G. }
 (Senior Member, IEEE) received the M.Sc. degree in mobile communications in 2008 and the Ph.D. degree in signal theory and communications in 2013, both from the Universitat Politècnica de Catalunya (UPC), Barcelona, Spain. From 2014 to 2017, he was a Postdoctoral Research Fellow with Aalto University, Espoo, Finland, and from 2017 to 2018, a Research Engineer with the Panasonic Research and Development Center, Germany. Since 2018, he has been with AUMOVIO Germany GmbH (formerly Continental Automotive Technologies, Germany), where he leads and contributes to research projects on wireless and vehicular communications (V2X), integrated sensing and communications (ISAC), and automotive connectivity for 5G-Advanced and 6G systems. He is also active in pre-standardization and standardization work as a delegate to 3GPP RAN1, task leader for ISAC work items of the 5G Automotive Association (5GAA), and rapporteur of Work Item 5 in Release 1 and Work Item 9 in Release 2 of the ETSI Industry Specification Group on ISAC (ETSI ISG ISAC).
\end{IEEEbiography}
\begin{IEEEbiography}[{\includegraphics[width=1in,height=1.25in,clip,keepaspectratio]{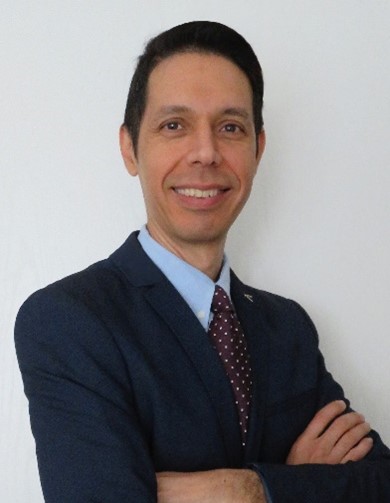}}]{Osvaldo Gonsa }
received the Ph.D. degree in electrical and computer engineering from Yokohama National University, Japan, in 1999, and the M.B.A. degree from the Kempten School of Business, Germany, in 2012. Since 1999 he has worked in research and standardization in the areas of core and radio access network. He is currently the Head of the Wireless Communications Technologies Group, Continental AG, Frankfurt, Germany. And since 2020 also serves as a member for the GSMA Advisory Board for automotive and the 6GKom Project of the German Federal Ministry of Education and Research.
\end{IEEEbiography}

\vspace{-2ex}\vfill\pagebreak

\end{document}